\documentclass{article}
\usepackage[top=30truemm,bottom=30truemm,left=25truemm,right=25truemm]{geometry}
\normalsize
\usepackage{multicol}

\usepackage{cite}  %
\usepackage{amsfonts}  %
\usepackage{set space} %
\usepackage{graphicx}
\usepackage{mathptmx}  
\usepackage{latexsym}
\usepackage{latexsym}
\usepackage{amsmath}
\usepackage{amsthm} 
\usepackage[all]{xy}

\title{
Toward construction of a consistent field theory with Poincar${\rm \acute{e}}$ covariance in terms of step-function-type basis functions showing confinement/deconfinement, mass-gap and Regge trajectory for non-pure/pure non-Abelian gauge fields
\thanks{
This paper is to appear in International Journal of Modern Physics A, Vol. 32 (2017) 1730017 DOI: 10.1142/S0217751X17300174 (title shortened).
} 
}

\author{
Kimichika Fukushima
\thanks{E-mail: kimichika1a.fukushima@glb.toshiba.co.jp; km.fukushima@mx2.ttcn.ne.jp 
Phone: +81-90-4602-0490 Phone/Fax: +81-45-831-8881}\\
Advanced Reactor System Engineering Department,\\
Toshiba Nuclear Engineering Service Corporation,\\
8, Shinsugita-cho, Isogo-ku, Yokohama 235-8523, Japan\\
 \\
Hikaru Sato\\
Emeritus, Department of Physics, Hyogo University of Education,\\
Yashiro-cho, Kato-shi, Hyogo 673-1494, Japan
}

\date{             }

\begin{document}

\maketitle

This article is a review by the authors concerning the construction of a Poincar${\rm \acute{e}}$ covariant (owing to spacetime continuum) field-theoretic formalism in terms of step-function-type basis functions without ultraviolet divergences. This formalism analytically derives confinement/deconfinement, mass-gap and Regge trajectory for non-Abelian gauge fields, and gives solutions for self-interacting scalar fields. Fields propagate in spacetime continuum and fields with finite degrees of freedom toward continuum limit have no ultraviolet divergence. Basis functions defined in a parameter spacetime are mapped to real spacetime. The authors derive a new solution comprised of classical fields as a vacuum and quantum fluctuations, leading to the linear potential between the particle and antiparticle from the Wilson loop. The Polyakov line gives finite binding energies and reveals the deconfining property at high temperatures. The quantum action yields positive mass from the classical fields and quantum fluctuations produces the Coulomb potential. Pure Yang-Mills fields show the same mass-gap owing to the particle-antiparticle pair creation. The Dirac equation under linear potential is analytically solved in this formalism, reproducing the principal properties of Regge trajectories at a quantum level. Further outlook mentions a possibility of the difference between conventional continuum and present wave functions responsible for the cosmological constant.

\par
\verb+ +
\par


\section{Introduction \label{sec:Sec1}}	

This article is a review concerning our published \cite{Fuku84,Fuku14,Fuku16} and reported \cite{Fuku85a} works about a quantum field-theoretic approach for non-Abelian Yang-Mills gauge fields \cite{YM54,Aber,PRam} using the localized basis functions with finite degrees of freedom in the spacetime continuum toward the continuum limit. After itemizing the motivations, we present a corresponding formalism and investigations on the properties of the non-Abelian gauge field, referring to sections where they are described.
\begin{enumerate}
\item
The relativistic quantum field theory \cite{Tomo46} has a limitation of ultraviolet divergences \cite{Weiss}, which stem from higher-order terms containing such as self-energies and vacuum polarizations \cite{Tomo48} due to the singularity of fields around a point source (particle). The construction of a consistent field theory without ultraviolet divergences is a requirement of fundamental theoretical physics.
Many attempts such as the lattice gauge theory \cite{KGW74a,KGW74b,Kog79,Kog83,Creu80,Creu83,Drou78,Drou83,APHa,HRot,Card,Baza,Call82,Call83,Call88,Wegn,Rebbi,CreuQ,Montv,Makee,Smit,DeGra06,DeGra10,Weisz} and (supersymmetric) string theory
\cite{SSt2,SSt3,SSt4} were proposed to remove the singularities. 
\item
A strong interacting system has specific properties such as color confinement \cite{Barg,Greens,WuHwa,Muta,Smil,Lake,Cahill} and mass-gap \cite{Luci,Chen} of the interaction field responsible for short-range interaction,
and its vacuum is investigated \cite{Debb94,MLer,Hoof76,Witt,Hoof79,Savv,Niel,Oles,Chod,Bela,Hoof76b,Enge,Debb97}.
Although the non-Abelian gauge field describes microscopic fundamental phenomena, efficient solutions and mechanism are not obtained using the basis equation \cite{Act79}. Moreover, explicit configurations of vector potentials in the scheme of the Yang-Mills theory are not clearly presented for the fundamental fermion confinement in the form of a particle-particle pair.
In a dual superconductor model \cite{GRip,MShi}, the static electric flux tube is squeezed in the superconductor and the interaction potential between a particle and an antiparticle becomes linear. However, the gauge waves, whose energy is lower than the superconducting gap, can pass through the superconductors \cite{CKit,MTin}.
\item
The masses of a pair of the fundamental particle and antiparticle were experimentally reported to be consistent with the Regge trajectory \cite{ColReg,Naga} in the form of the squares of the masses proportional to the total angular momentum.
A classical mechanical approach shows that the Hamiltonian is comprised of a linear potential and repulsive rotational energy \cite{Miya,Maki}, whose essence is summarized in Eqs. (\ref{eqn:EMFM})-(\ref{eqn:EMTO}) of Subsection \ref{sec:Sec61}. The classical Hamiltonian is not related to the Dirac equation under the linear potential produced by the non-Abelian Yang-Mills fields. At the quantum level, the principal whole Regge trajectory is not sufficiently reproduced by the other theoretical/numerical calculations \cite{Kabu,AbeF,Tez91,Tez13}. 
Moreover, the electric field is not squeezed above the superconducting critical temperature. The bound particle and antiparticle will freely move at high temperatures.
\cite{Bohr,Zajc,Plmer1,Plmer2,Satz,Ander11,Ander10}
Experiments at high energies (temperatures) reported that particles and antiparticles resemble a fluid without viscosity.
\end{enumerate}
\par
Considering the above motivations, the contents of this article are organized as follows.
\begin{itemize}
\item
In Section \ref{sec:Sec2},
we construct a consistent field theory based on the finite element theory (method) (FEM) \cite{Fuku84,Fuku14,Fuku16,FemMW} for the first motivation above, but it rather differs from that proposed by Bender {\it et al.} \cite{BenMS} owing to the difference in motivations. The finite element theory expresses the wave function in terms of localized basis functions with finite degrees of freedom in the spacetime continuum. The final results calculated using this method are obtained in the continuum limit. This formalism enables the use of differentiation, which is consistent with the fields propagating in the spacetime continuum. The suppression of the oscillation of the wave function by the step-function-type basis function and the finite degrees of freedom remove the divergence of field quantities. The reason for the step-function-type basis functions being employed in the present theory is that the gauge transformation involving the non-Abelian gauge field requires cancellation not only for the variation of a fermion phase but also for the product of the variation of the fermion phase and the non-Abelian gauge field. Furthermore, these basis functions localized in spacetime are defined in a parameter spacetime continuum and mapped to the real spacetime continuum. By regarding the basis function like a physical object in the spacetime continuum, the formalism becomes Poincar${\rm \acute{e}}$ covariant owing to the property of continuum spacetime. This field theory is applied to a self-interacting scalar field in low-dimensions by using the path integral and variational calculus.
\item
Section \ref{sec:Sec3} summaries the non-Abelian Yang-Mills theory with such as matrix representation of Lie algebra, gauge invariance and the related Wilson loop as a preliminary for subsequent sections.
\item
Section \ref{sec:Sec4} explicitly show an example of analytic classical vector potentials \cite{Fuku14}, comprised of a localized function and an unlocalized function to describe a confined fundamental fermion-antifermion pair in the center-of-mass frame, corresponding to the aforementioned second motivation.
It is shown that the total classical field, shifted from the zero-amplitude field, is a solution of the classical non-Abelian Yang-Mills equations of motion as a non-perturbative vacuum. The localized and unlocalized functions are quite different types of functions, since the localized function has a soliton-like shape of a thin sheet in spacetime, whereas the unlocalized function is a spherical wave function. These two functions have useful properties such that the localized function of the classical field configuration leads to a confining potential derived from the classical Wilson loop, whereas the unlocalized function gives no contribution to the Wilson loop. The confinement is caused by the trace of matrix polynomials in Lie algebra, which does not appear in the case of the Abelian gauge field. The existence of this classical configuration in the confining phase, which satisfies the classical equations of motion, is verified by the energy lowering in the Wilson loop, compared to the configuration in the Coulomb phase. The classical vector potentials, we have found, have not been mentioned anywhere in literature \cite{Act79}.
\item
In Section \ref{sec:Sec5}, it is shown that the present formalism, in which the fields are expressed in terms of step-function-type basis functions for non-Abelian Yang-Mills fields, is gauge invariant \cite{Fuku14}. Quantum fields, which are fluctuations around the classical field presented in Section \ref{sec:Sec4} as a vacuum, are expressed in terms of step-function-type basis functions in the path integral. The quadratic terms are analytically diagonalized, and eigenvalues take positive values under a non-periodic boundary condition to avoid zero eigenvalue caused by a periodic boundary condition. The quantum fluctuation for the small coupling constant according to the asymptotic freedom leads to the Coulomb potential, and the total potential derived is the sum of the classical linear potential and the quantum Coulomb potential.
\item
We note that the classical field produces the non-zero positive mass term for quantum fluctuations of the non-Abelian Yang-Mills field in the action. This phenomenon also occurs for the pure non-Abelian Yang-Mills fields, because a particle-antiparticle pair of the gauge field is created. The confined particle and antiparticle of the gauge source require the energy (mass) for the deconfinement \cite{Polya}. The present formalism given in this article is Poincar${\rm \acute{e}}$ invariant with a cut-off to avoid ultraviolet divergences as mentioned in Section \ref{sec:Sec2} and the gauge-invariance as described in this section, which states that this formalism demonstrates the existence of the non-Abelian Yang-Mills fields.
Furthermore, it was shown that the pure non-Abelian Yang-Mills fields have the mass-gap. Consequently, the present formalism offers a solution to the requirement of the fundamental field theory and questions by Pauli.
\item
Section \ref{sec:Sec6} presents the analytical solutions of the Dirac equation under a confining linear potential \cite{Fuku16}, considering  the aforementioned third motivation.
We obtain the eigenenergies of a confined fundamental fermion-antifermion pair using the formalism \cite{Fuku84,Fuku14} in terms of the step-function-type basis functions with finite degrees of freedom in the spacetime continuum, mentioned in Section \ref{sec:Sec2}. The present formalism enables an analytical calculation unlike numerical computer simulations. The total Hamiltonian involving a given linear potential leads to the Dirac equation in spherical coordinates by variational calculus. The secular equation in the Hamiltonian matrix form is diagonalized analytically, for the large rotational energy compared to the constituent particle masses, corresponding to the classical mechanical Hamiltonian. The lowest eigenvalue derived is a function of the string tension and the relativistic quantum number of the total angular momentum \cite{Schf,Dira}. We emphasize that the classical mechanical Hamiltonian does not
contain
the relativistic quantum number for the total angular momentum. The squared energies (masses),
reproduce
the principal properties of the Regge trajectory \cite{ColReg} at the quantum level.
\item
In Section \ref{sec:Sec7} (Further outlook), it is stated that the difference between the energy derived by the continuum gravitational theory and that derived by the aforementioned step-function-type basis functions might reveal the cosmological constant of the order of the matter (atoms) energy. The conventional continuum gravitational theory cannot strictly describe the energy at the
Planck
scale because of the ultraviolet divergence, whereas the step-function-type basis functions describe the phenomena more precisely. Then, the aforementioned energy difference
due to the difference in the wave function of fields expressing such as curvatures
would compensate the energy of the continuum theory by involving the aforementioned energy difference into the cosmological constant (renormalizing the cosmological constant in an extended meaning) of the gravitational Einstein equation.
The order of magnitude of this energy difference would be the same order of magnitude of the matter (atoms).
\item
In Section \ref{sec:Sec8}, we summarize the  conclusions.
\end{itemize}
\par

\section{Poincar${\rm \acute{e}}$-covariant formalism for fields with finite degrees of freedom in real spacetime associated with a parameter spacetime \label{sec:Sec2}}

\subsection{Parameter spacetime and map to real spacetime \label{sec:Sec21}}

We first construct a formulation of fields that propagate in continuum spacetime with finite degrees of freedom. It is shown that, owing to the spacetime continuity, this theoretical scheme in the Lagrangian form is covariant under the Poincar${\rm \acute{e}}$ transformation including the Lorentz transformation.
In this article, a four-dimensional Minkowski time-space point is denoted as $x=(x_{0},x_{1},x_{2},x_{3})=(t,{\bf x})$ ($x=x_{0}$  represents time and ${\bf x}=(x_{1},x_{2},x_{3}$) represents space) with the velocity of light set to be unity as $c=1$. We define the real norm (metric) squared and four-dimensional volume element in the Minkowski time-space as
\begin{eqnarray}
x_{\mu}x_{\mu}
=-x_{0}x_{0}+x_{i}x_{i}
=-x_{0}x_{0}+x_{1}x_{1}+x_{2}x_{2}+x_{3}x_{3},
\end{eqnarray}
\begin{eqnarray}
dx^{4}=dx_{0}dx_{1}dx_{2}dx_{3},
\end{eqnarray}
respectively (Latin space index $i$ runs from $1$ to $3$).
In the Euclidean time-space case, the above quantity is written as
\begin{eqnarray}
x_{\mu}x_{\mu}
=x_{0}x_{0}+x_{i}x_{i}
=x_{0}x_{0}+x_{1}x_{1}+x_{2}x_{2}+x_{3}x_{3}.
\end{eqnarray}
\par
Further, we introduce a four-dimensional parameter spacetime with coordinates $x_{{\rm P}\mu}$, which is divided into hypercubes. 
These hypercubes in the parameter spacetime are mapped to hyper-octahedrons in real spacetime, whose shape of (hyper-) surfaces is arbitrary.
The coordinates of the parameter time-space are expressed as $(x_{\rm P 0},x_{\rm P 1},x_{\rm P 2},x_{\rm P 3})$$=(t_{\rm P},x_{\rm P},y_{\rm P},z_{\rm P})$ and a lattice (grid) point in the parameter spacetime is defined as $x_{{\rm P} p}=x_{{\rm P} (k,l,m,n)}=(t_{{\rm P}(k)},x_{{\rm P}(l)},y_{{\rm P}(m)},z_{{\rm P}(n)})$, where the subscript $p$ denotes $p=(k,l,m,n)$, and the indices $k, l, m$ and $n$ run from $1$ to $N_{{\rm P}\mu}$ (for $\mu=0,1,2$ and $3$), respectively.
Figure 1 shows a simple example of a two-dimensional $(t_{{\rm P}},x_{\rm P})$-plane in the four-dimensional parameter spacetime. The corners (grid points) of the hypercube are denoted by $P_{kl}$, which is located at $(t_{{\rm P}(k)},x_{{\rm P}(l)})$ in the plane (more precisely $(t_{{\rm P}(k)},x_{{\rm P}(l)},0,0)$), where integer indices $k$ and $l$ (in the sample case) run from 1 to $N_{{\rm P}t}=3$ and from 1 to $N_{{\rm P}x}=3$, respectively. The corners of the hypercube are denoted such as $P_{11}$, $P_{12}$, $P_{22}$ and $P_{21}$.
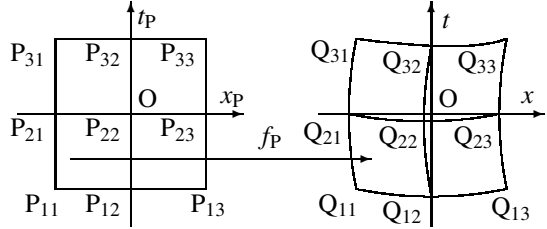
\begin{figure}[tbh]
\begin{center}
\setlength{\unitlength}{1 mm}
\begin{picture}(80,50)(-40,-25)
\put( -35,   0){\vector(1,0){30}}
\put( -20,-15){\vector(0,1){30}}
\put( -30, 10){\line(1,0){20}}
\put( -30,-10){\line(1,0){20}}
\put( -30,-10){\line(0,1){20}}
\put( -10,-10){\line(0,1){20}}
\put( -19,  1){${\rm O}$}
\put( -19,  12){${t_{\rm P}}$}
\put( - 8,  1.5){${x_{\rm P}}$}
\put(  -34, -13){${\rm P_{11}}$}
\put(  -26, -13){${\rm P_{12}}$}
\put(  -12, -13){${\rm P_{13}}$}
\put(  -36, -3){${\rm P_{21}}$}
\put(  -26, -3){${\rm P_{22}}$}
\put(  -16, -3){${\rm P_{23}}$}
\put(  -36, 7){${\rm P_{31}}$}
\put(  -26, 7){${\rm P_{32}}$}
\put(  -16, 7){${\rm P_{33}}$}
\put(   5,   0){\vector(1,0){30}}
\put(  20, -15){\vector(0,1){30}}
\qbezier(10, 10)(25,  8)(30, 10)
\qbezier( 9.01,0)(19.01,-2.)(29.01,0)
\qbezier(10,-10)(20,-12)(30,-10)
\qbezier(10,-10)(8, 0)(10, 10)
\qbezier(20,-10.99)(18,-0.99)(20,9.01)
\qbezier(30,-10)(28, 0)(30, 10)
\put(  21,  1){${\rm O}$}
\put(  21.5,  11.5){${t}$}
\put(  32,  1.5){${x}$}
\put(   5, -13){${\rm Q_{11}}$}
\put(   13.5, -14){${\rm Q_{12}}$}
\put(   28, -13.5){${\rm Q_{13}}$}
\put(   3, -3.5){${\rm Q_{21}}$}
\put(   13, -4){${\rm Q_{22}}$}
\put(   23, -4){${\rm Q_{23}}$}
\put(   4, 7.5){${\rm Q_{31}}$}
\put(   13.5, 6){${\rm Q_{32}}$}
\put(   23.5, 6){${\rm Q_{33}}$}
\put(  -28,-6){\vector(1,0){40}}
\put(  -3, -4){${f_{{\rm P}}}$}
\end{picture}
\caption{
Two-dimensional $(t_{\rm P},x_{\rm P})$-plane
(with time and space coordinates $t_{\rm P}$ and $x_{\rm P}$, respectively)
in a four-dimensional parameter spacetime composed of hypercubes on the left-hand side, mapped by a (total) function 
$f_{\rm P}$ (with four components $f_{{\rm P}\mu}$)
to the $(t,x)$-plane
in four-dimensional real spacetime on the right. The position of the point 
$P_{kl}$ on the left is $(t_{{\rm P}(k)},x_{{\rm P}(l)})$ (more precisely, 
$(t_{{\rm P}(k)},x_{{\rm P}(l)},0,0)$ in 
$(t_{\rm P},x_{\rm P},y_{\rm P},z_{\rm P})$ coordinates), 
and the corresponding related point on the right is $Q_{kl}$
 located at $(t_{(k)},x_{(l)})$ (more precisely,
$(t_{(k)},x_{(l)},0,0)$ in $(t,x,y,z)$ coordinates).
}
\label{fig:fig1}
\end{center}
\end{figure}
\par
This parameter space is mapped to the four-dimensional real spacetime using the following total function
\begin{eqnarray}
\label{eqn:DfPMP}
x=f_{\rm P}(x_{\rm P}),
\end{eqnarray}
whose components are
\begin{eqnarray}
\label{eqn:mapx}
x_{\mu}=f_{{\rm P}\mu}(x_{\rm P})
=f_{{\rm P}\mu}(t_{{\rm P}},{\bf x}_{{\rm P}}).
\end{eqnarray}
The total inverse map is written as
\begin{eqnarray}
x_{\rm P}=f_{\rm R}(x)=f_{\rm P}^{-1}(x),
\end{eqnarray}
with components
\begin{eqnarray}
x_{{\rm P}\mu}=f_{{\rm R}\mu}(x)=f_{{\rm R}\mu}(t,{\bf x}).
\end{eqnarray}
In Fig. 1, each grid point $P_{kl}$ in the parameter spacetime is mapped to $Q_{kl}$ in real spacetime.
More generally, the time coordinate $t_{\rm P}$ in the parameter spacetime at a three-dimensional hypersurface $\sigma_{{\rm P}t}$, which is normal to the time axis (such as in the Euclidean spacetime) is mapped to time $t_{\sigma_{t}}$ in real spacetime at a spacelike hypersurface $\sigma_{t}$ using the following function of the Fourier series form (considering the notations such as $n_i f_{{\rm P}i}=n_1 f_{{\rm P}1} + n_2 f_{{\rm P}2} + n_3 f_{{\rm P}3}$)
\begin{eqnarray}
\label{eqn:HSt1}
t_{\sigma_{t}}=x_{0\sigma_{x_0}}=
\sum_{n_1,n_2,n_3}
[ a_{n_1,n_2,n_3}\sin(\frac{2\pi n_i}{L_{\rm S}} x_i)
 +b_{n_1,n_2,n_3}\cos(\frac{2\pi n_i}{L_{\rm S}} x_i) ],
\end{eqnarray}
where $n_i$ ($i=1, 2, 3$) are non-negative integers. The symbols $a_{n_1,n_2,n_3}$ and $b_{n_1,n_2,n_3}$ are arbitrary constants, and $L_{\rm S}$ is a large positive length for the Fourier series.
From Eq. (\ref{eqn:mapx}) and Eq. (\ref{eqn:HSt1}), we find
\begin{eqnarray}
\label{eqn:HSta}
t_{\sigma_{t}}=
\sum_{n_1,n_2,n_3}
[ a_{n_1,n_2,n_3}\sin(\frac{2\pi n_i}{L_{\rm S}} f_{{\rm P}i})
 +b_{n_1,n_2,n_3}\cos(\frac{2\pi n_i}{L_{\rm S}} f_{{\rm P}i}) ].
\end{eqnarray}
By denoting the increments in the $x_{\mu}$ directions as
$\Delta
x_{\mu}
$, 
the spacelike hypersurface in the Minkowski spacetime satisfies
\begin{eqnarray}
(\Delta t)^2 < (\Delta
x
_i)^2 .
\end{eqnarray}
\par
Similarly,
the $x_{\rm P}$-coordinate in the parameter spacetime at a three-dimensional hypersurface $\sigma_{{\rm P}x}$, which is normal to the $x_{\rm P}$-axis, is mapped to the $x$-coordinate $x_{\sigma_{t}}$ in real spacetime at a hypersurface $\sigma_{x}$  using the following function
\begin{eqnarray}
\nonumber
x_{\sigma_{x}}=
\sum_{n_0,n_2,n_3} [ a_{n_0,n_2,n_3} 
\sin(- \frac{2 \pi n_0}{L_{\rm S}}x_0 + \frac{2 \pi n_2}{L_{\rm S}}x_2
     + \frac{2 \pi n_3}{L_{\rm S}}x_3)
\end{eqnarray}
\begin{eqnarray}
+b_{n_0,n_2,n_3}
\cos(- \frac{2 \pi n_0}{L_{\rm S}}x_0 + \frac{2 \pi n_2}{L_{\rm S}}x_2
     + \frac{2 \pi n_3}{L_{\rm S}}x_3) ],
\end{eqnarray}
where the symbol $n_0$ stands for non-negative integer. Hence, the hypercubes in the parameter spacetime are mapped to the hyper-octahedrons with arbitrary hypersurface shapes in the real spacetime.
\par

\subsection{Step-function-type basis functions localized in spacetime continuum 
\label{sec:Sec22}}

Our purpose is to construct a field theory, considering that fields propagate in the spacetime continuum and the fields are well defined when the degrees of freedom are finite. In order to realize the formulation of the field theory, fields are expressed in terms of step-function-type basis functions localized in a parameter spacetime continuum introduced in the previous Subsection \ref{sec:Sec21}, and the fields are mapped to the real spacetime continuum. 
For coordinates of the parameter spacetime $(t_{\rm P},x_{\rm P},y_{\rm P},z_{\rm P})=(x_{\rm P 0},x_{\rm P 1},x_{\rm P 2},x_{\rm P 3})$, and lattice (grid) points at the corners of divided hypercubes $x_{{\rm P} p}=x_{{\rm P} (k,l,m,n)}=(t_{{\rm P}(k)},x_{{\rm P}(l)},y_{{\rm P}(m)},z_{{\rm P}(n)})$ (the indices $p=(k,l,m,n)$ and $k, l, m$ and $n$ run over integers from $1$ to $N_{{\rm P}\mu}$ with $\mu=0,1,2,3$), let $\Delta$ be an infinitesimal lattice spacing. Then, we define the points between lattice points as $t_{{\rm P}(k - 1/2)}=t_{{\rm P}(k)} - \Delta/2$ and $t_{{\rm P}(k + 1/2)}=t_{{\rm P}(k)} + \Delta/2$ in the case of time coordinate in the parameter spacetime.
\par
We introduce the following smooth function of $t_{\rm P}$ from $-\epsilon_{\rm G}$ to $+\epsilon_{\rm G}$
\begin{eqnarray}
\label{eqn:fGINT}
f_{\rm G}(t_{\rm P})=
\frac{
\int_{-\epsilon_{\rm G} }^{ t_{P} }
d t_{\rm P}^{\prime}(\frac{\pi}{a_{\rm G}})
\exp[-a_{\rm G}(t_{\rm P}^{\prime})^2]
}
{
\int_{ -\epsilon_{\rm G} }^{+\epsilon_{\rm G} }
d t_{\rm P}^{\prime}(\frac{\pi}{a_{\rm G}})
\exp[-a_{\rm G}(t_{\rm P}^{\prime})^2]
},
\end{eqnarray}
where $a_{\rm G}$ is a positive constant, and $\epsilon_{\rm G}$ is an infinitesimal positive quantity that is taken to be $\epsilon_{\rm G} \rightarrow 0$ after calculations. The above function is a normalized integral of the Gauss function, and takes from 0 to 1.
The derivative of the above function becomes
\begin{eqnarray}
\frac{\partial f_{\rm G}(t_{\rm P})}{\partial t_{P}}=\delta(t_{P})
\hspace{1ex}\mbox{ in the limit of }a_{\rm G} \rightarrow \infty,
\end{eqnarray}
where $\delta(t_{P})$ is the Dirac's delta function.
Using the integral in Eq. (\ref{eqn:fGINT}), we define a step-function such as a smooth basis function,
\begin{eqnarray}
\label{Eq:EGDF}
\hspace{-4ex}
\Omega^{\rm {\tilde{\rm E}G}}_{k}(t_{\rm P})=
\left\{\begin{array}{ll}
0 & \mbox{ for 
$t_{\rm P} \leq t_{{\rm P}(k-1/2)} -\epsilon_{\rm G}$ }\\
\mbox{ } & \mbox{ } \\
f_{\rm G}(t_{\rm P}+t_{{\rm P}(k-1/2)}) & \mbox{ for 
$t_{{\rm P}(k-1/2)}-\epsilon_{{\rm G}} <t_{\rm P} < t_{{\rm P}(k-1/2)}+\epsilon_{{\rm G}}$}\\
\mbox{ } & \mbox{ } \\
1 & \mbox{ for 
$t_{{\rm P}(k-1/2)}+\epsilon_{{\rm G}} \leq t_{\rm P} < t_{{\rm P}(k+1/2)}-\epsilon_{{\rm G}}$}\\
\mbox{ } & \mbox{ } \\
0 & \mbox{for 
$t_{{\rm P}(k+1/2)} -\epsilon_{\rm G} \leq t_{\rm P}$
}
\end{array}\right. .
\end{eqnarray}
The above basis function $\Omega^{\rm \tilde{E}G}_{k}(t_{\rm P})$ takes positive values in the region between $t_{{\rm P}(k-1/2)}-\epsilon_{{\rm G}}$ and $t_{{\rm P}(k+1/2)}-\epsilon_{{\rm G}}$ around the point $t_{{\rm P}(k)}$. This function increases from 0 to 1, subsequently maintains the value of 1, and decreases to 0 with the increase of $t_{\rm P}$.
\par
For the sake of convenience, we define the simpler basis function as
\begin{eqnarray}
 \label{eqn:DeEB}
\Omega^{\tilde{\rm E}}_{k}(t_{\rm P})=
\left\{\begin{array}{ll}
0 & \mbox{ for 
$t_{\rm P} < t_{{\rm P}(k-1/2)}$  }\\
\mbox{ } & \mbox{ } \\
1 & \mbox{ for 
$t_{{\rm P}(k-1/2)} \leq t_{\rm P} < t_{{\rm P}(k+1/2)}$ }\\
\mbox{ } & \mbox{ } \\
0 & \mbox{for 
$t_{{\rm P}(k+1/2)} \leq t_{\rm P}$ }
\end{array}\right. ,
\end{eqnarray}
which plays an important role in this article. 
Subsequently, the derivative is given by
\begin{eqnarray}
 \label{eqn:DeDE}
\frac{d \Omega^{\tilde{\rm E}}_{k}(t_{\rm P})}{d t_{\rm P}}
=
\left\{\begin{array}{ll}
0
& \mbox{ for $t_{\rm P} < t_{{\rm P}(k-1/2)}$ }\\
\mbox{ } & \mbox{ } \\
\delta(t_{\rm P}-t_{{\rm P}(k-1/2)})
& \mbox{ at $t_{\rm P}=t_{{\rm P}(k-1/2)}$}\\
\mbox{ } & \mbox{ } \\
0
& 
\mbox{ for $t_{{\rm P}(k-1/2)} <t_{\rm P} < t_{{\rm P}(k+1/2)}$}\\
\mbox{ } & \mbox{ } \\
-\delta(t_{\rm P}-t_{{\rm P}(k+1/2)})
& \mbox{ at $t_{\rm P} = t_{{\rm P}(k+1/2)}$}\\
\mbox{ } & \mbox{ } \\
0
& \mbox{ for $t_{\rm P} > t_{{\rm P}(k+1/2)}$ }
\end{array}\right. .
\end{eqnarray}
In the four-dimensional case, the basis functions $\Omega^{4\tilde{\rm E}}_{p}(x_{\rm P})$ and $\Omega^{\delta 3\tilde{\rm E}}_{\mu p}(x_{\rm P})$ are defined as
\begin{eqnarray}
 \label{eqn:De4EB}
 \Omega^{4\tilde{\rm E}}_{p}(x_{\rm P})
=\Omega^{4\tilde{\rm E}}_{(k,l,m,n)}(x_{\rm P})
=\Omega^{\tilde{\rm E}}_{k}(t_{
\rm
P
})\Omega^{\tilde{\rm E}}_{l}(x_{
\rm
P
})
\Omega^{\tilde{\rm E}}_{m}(y_{
\rm
P
})\Omega^{\tilde{\rm E}}_{n}(z_{
\rm
P
}),
\end{eqnarray}
\begin{eqnarray}
 \label{eqn:DeD4E}
\nonumber
 \Omega^{\delta 3\tilde{\rm E}}_{\mu p}(x_{\rm P})
=\frac{d \Omega^{4\tilde{\rm E}}_{p}(x_{\rm P})}{d x_{\mu {\rm P}}}\Delta
=\frac{d \Omega^{4\tilde{\rm E}}_{(k,l,m,n)}(x_{\rm P})}
{d x_{\mu {\rm P}}}\Delta
\end{eqnarray}
\begin{eqnarray}
=\Omega^{\delta 3\tilde{\rm E}-}_{\mu p}(x_{\rm P})
-\Omega^{\delta 3\tilde{\rm E}+}_{\mu p}(x_{\rm P}).
\end{eqnarray}
Here, the above basis functions
$\Omega^{\delta 3\tilde{\rm E}-}_{\mu p}(x_{\rm P})$
and
$\Omega^{\delta 3\tilde{\rm E}+}_{\mu p}(x_{\rm P})$
with non-vanishing values
for $\mu=0$ are written as
\begin{eqnarray}
\label{eqn:DB3N}
\nonumber
\Omega^{\delta 3{\tilde{\rm E}}-}_{0 p}(t_{\rm P}=t_{{\rm P}(k-1/2)})
=\frac{d \Omega^{4\tilde{\rm E}}_{p}(x_{\rm P})}{d t_{\rm P}}
|_{ t_{\rm P}=t_{{\rm P}(k-1/2)} }
\Delta
\end{eqnarray}
\begin{eqnarray}
\nonumber
=\frac{d \Omega^{\tilde{\rm E}}_{k}(t_{\rm P})}
 {d t_{\rm P}}
|_{t_{\rm P}=t_{{\rm P}(k-1/2)}}
\Omega^{\tilde{\rm E}}_{l}(x_{
\rm
P
})\Omega^{\tilde{\rm E}}_{m}(y_{
\rm
P
})
\Omega^{\tilde{\rm E}}_{n}(z_{
\rm
P
})\Delta
\end{eqnarray}
\begin{eqnarray}
=\delta(t_{\rm P}-t_{{\rm P}(k-1/2)})
\Omega^{\tilde{\rm E}}_{l}(x_{
\rm
P
})\Omega^{\tilde{\rm E}}_{m}(y_{
\rm
P
})
\Omega^{\tilde{\rm E}}_{n}(z_{
\rm
P
})\Delta,
\end{eqnarray}
\begin{eqnarray}
\nonumber
\label{eqn:DB3P}
\Omega^{\delta 3{\tilde{\rm E}}+}_{0 p}(t_{\rm P}=t_{{\rm P}(k+1/2)})
=-\frac{d \Omega^{4\tilde{\rm E}}_{p}(x_{\rm P})}{d t_{\rm P}}
|_{t_{\rm P}=t_{{\rm P}(k+1/2)}}
\Delta
\end{eqnarray}
\begin{eqnarray}
\nonumber
=-\frac{d \Omega^{\tilde{\rm E}}_{k}(
t_{\rm P}
)}
  {d t_{\rm P}}
|_{t_{\rm P}=t_{{\rm P}(k+1/2)}}
\Omega^{\tilde{\rm E}}_{l}(x_{
\rm
P
})\Omega^{\tilde{\rm E}}_{m}(y_{
\rm
P
})
\Omega^{\tilde{\rm E}}_{n}(z_{
\rm
P
})\Delta
\end{eqnarray}
\begin{eqnarray}
=\delta(t_{\rm P}-t_{{\rm P}(k+1/2)})
\Omega^{\tilde{\rm E}}_{l}(x_{
\rm
P
})\Omega^{\tilde{\rm E}}_{m}(y_{
\rm
P
})
\Omega^{\tilde{\rm E}}_{n}(z_{
\rm
P
})\Delta.
\end{eqnarray}
\par
The above basis functions have the following properties:
\begin{eqnarray}
\int dt_{\rm P} \Omega^{\tilde{\rm E}}_{k}(t_{\rm P})=\Delta,
\end{eqnarray}\begin{eqnarray}
\Omega^{\tilde{\rm E}}_{k}(t_{\rm P})
\Omega^{\tilde{\rm E}}_{k^{\prime}}(t_{\rm P})
=\delta_{k,k^{\prime}}\Omega^{\tilde{\rm E}}_{k}(t_{\rm P}),
\end{eqnarray}
where $\delta_{k,k^{\prime}}$ is Kroneker's delta.
\par
Furthermore, the basis functions in real spacetime can be defined, by mapping the basis functions in the parameter space to real spacetime, as
\begin{eqnarray}
\label{eqn:DfBSR1}
 \Omega^{4\tilde{\rm E}}_{{\rm R}p}(x)
=\Omega^{4\tilde{\rm E}}_{p}(f_{\rm P}(x_{\rm P})),
\end{eqnarray}
\begin{eqnarray}
\label{eqn:DfBSR2}
 \Omega^{\delta 3\tilde{\rm E}}_{\rm R{\mu p}}(x)
=\Omega^{\delta 3\tilde{\rm E}}_{\mu p}(f_{\rm P}(x_{\rm P})).
\end{eqnarray}

\subsection{Poincar${\rm \acute{e}}$ transformation of fields described in terms of basis functions in spacetime continuum \label{sec:Sec23}}

In this subsection, we present Poincar${\rm \acute{e}}$-covariant field formalism with finite degrees of freedom. We note that the hyper-octahedrons in spacetime may be regarded as if they are physical objects, and the Poincar${\rm \acute{e}}$ covariance is due to the spacetime continuum (with the existence of derivatives). Fields are expressed in terms of real spacetime basis functions, which have the shape of step-functions in a parameter space and are mapped to real spacetime, as defined in the previous Subsection \ref{sec:Sec22}.
We consider an example of a self-interacting scalar field $\phi(x)$ with the action
\begin{eqnarray}
S_{3}=S_{3}^{(2)}+S_{3}^{(3)},
\end{eqnarray}
where
\begin{eqnarray}
\nonumber
S_{3}^{(2)}=
\int dx^4 
\{-\frac{1}{2}
 \frac{\partial \phi(x)}{\partial x_{\mu}}
 \frac{\partial \phi(x)}{\partial x_{\mu}}
-\frac{1}{2}m^2[\phi(x)]^2
\}
\end{eqnarray}
\begin{eqnarray}
=\int dx^4 
\{-\frac{1}{2}
 \partial_{\mu} \phi(x)
 \partial_{\mu} \phi(x)
-\frac{1}{2}m^2[\phi(x)]^2
\},
\end{eqnarray}
\begin{eqnarray}
S_{3}^{(3)}=\int dx^4 
\{
-\frac{g_3}{3!} [\phi(x)]^3
\},
\end{eqnarray}
and $m$ stands for the mass, with $g_3$ being a coupling constant.
\par
Further, the field is expanded in terms of the basis functions given in Eq. (\ref{eqn:DfBSR1}) as the following detailed form
\begin{eqnarray}
\nonumber
\phi(x)
=\sum_p \phi_p \Omega^{4\tilde{\rm E}}_{{\rm R}p} (f_{{\rm P}}(x_{\rm P}))
\end{eqnarray}
\begin{eqnarray}
\nonumber
=\sum_p \phi_p \Omega^{4\tilde{\rm E}}_{{\rm R}p}
(f_{{\rm P}t}(t_{\rm P},{\bf x}_{\rm P}
),f
_{{\rm P}x}(t_{\rm P},{\bf x}_{\rm P}
),f
_{{\rm P}y}(t_{\rm P},{\bf x}_{\rm P}
),f
_{{\rm P}z}(t_{\rm P},{\bf x}_{\rm P}))
\end{eqnarray}
\begin{eqnarray}
\nonumber
=\sum_p \phi_p \Omega^{4\tilde{\rm E}}_{{\rm R}p}
(f_{{\rm P}t}(x_{\rm P}
),f
_{{\rm P}x}(x_{\rm P}
),f
_{{\rm P}y}(x_{\rm P}
),f
_{{\rm P}z}(x_{\rm P}))
\end{eqnarray}
\begin{eqnarray}
=\sum_p \phi_p \Omega^{4\tilde{\rm E}}_{{\rm R}p}(t, {\bf x})
=\sum_p \phi_p \Omega^{4\tilde{\rm E}}_{{\rm R}p}(x).
\end{eqnarray}
Then, each action term is expressed as
\begin{eqnarray}
\label{eqn:ac32}
\nonumber
S_{3}^{(2)}=\int dx^4 \{-\frac{1}{2}
\sum_{p,q}\phi_p\phi_q
\partial_{\mu}  \Omega^{4\tilde{\rm E}}_{{\rm R}p}(x)
\partial_{\mu}  \Omega^{4\tilde{\rm E}}_{{\rm R}q}(x)
\end{eqnarray}
\begin{eqnarray}
-\frac{1}{2}m^2\sum_{p,q}\phi_p\phi_{q}
\Omega^{4\tilde{\rm E}}_{{\rm R}p}(x) \Omega^{4\tilde{\rm E}}_{{\rm R}q}(x)
\},
\end{eqnarray}
\begin{eqnarray}
\label{eqn:ac33}
S_{3}^{(3)}=\int dx^4 
\{-\frac{g_3}{3!}
\sum_{
p,q,r
}
\phi_p\phi_q\phi_r
\Omega^{4\tilde{\rm E}}_{{\rm R}p}(x) \Omega^{\rm 4
\tilde{\rm E}
}_{{\rm R}q}(x)
\Omega^{4\tilde{\rm E}}_{{\rm R}r}(x)
\}.
\end{eqnarray}
\par
Further, we consider the Poincar${\rm \acute{e}}$ transformation
\begin{eqnarray}
\label{eqn:Poin1}
x_{{\rm L} \mu}=f_{{\rm L} \mu}(x_{\nu})
=\Lambda_{\mu \nu} x_{\nu} +a_{\mu},
\end{eqnarray}
which is expressed in short as
\begin{eqnarray}
\label{eqn:Poin2}
x_{\rm L}=f_{\rm L}(x)=\Lambda x+a,
\end{eqnarray}
with its inverse transformation
\begin{eqnarray}
\label{eqn:Poin3}
x=f_{\rm L}^{-1}(x_{\rm L})
=
\Lambda^{-1}(x_{\rm L}-a).
\end{eqnarray}
Using Eq. (\ref{eqn:Poin3}), we obtain
\begin{eqnarray}
\label{eqn:phiL}
\phi(x)
=\phi(f_{\rm L}^{-1}(x_{\rm L}))
=\phi(
\Lambda^{-1}(x_{\rm L}-a)
).
\end{eqnarray}
Then, the field, which is expressed by $\phi(x)$ in the frame with $x$ coordinates, is defined, in the frame with $x_{\rm L}$ coordinates, that
\begin{eqnarray}
\label{eqn:DphiL}
\phi_{\rm L}(x_{\rm L})
=\phi(x).
\end{eqnarray}
Similarly, for the basis functions given by Eq. (\ref{eqn:DfBSR1}) in real spacetime, we have the following relation, corresponding to Eq. (\ref{eqn:phiL}), 
\begin{eqnarray}
\Omega^{4\tilde{\rm E}}_{{\rm R}p}(x)
=\Omega^{4\tilde{\rm E}}_{{\rm R}p}(f_{\rm L}^{-1}(x_{\rm L}))
=\Omega^{4\tilde{\rm E}}_{{\rm R}p}(
\Lambda^{-1}(x_{\rm L}-a)
).
\end{eqnarray}
Hence, analogous to Eq. (\ref{eqn:DphiL}), we define that
\begin{eqnarray}
\label{eqn:basL}
\Omega^{4\tilde{\rm E}}_{{\rm L}p}(x_{\rm L})
=\Omega^{4\tilde{\rm E}}_{{\rm R}p}(x).
\end{eqnarray}
\par
Using Eqs. (\ref{eqn:DphiL}) and (\ref{eqn:basL}), the comprising terms of the action in Eqs. (\ref{eqn:ac32}) and (\ref{eqn:ac33}) are transformed to
\begin{eqnarray}
\nonumber
S_{3}^{(2)}=
\int dx_{\rm L}^4 J_{f_{\rm L}}
\{-\frac{1}{2}
\sum_{p,q}\phi_p\phi_q
 \frac{\partial x_{\rm L \nu}}{\partial x_{\mu}}
 \frac{\partial \Omega^{4\tilde{\rm E}}_{{\rm L}p}(x_{\rm L})}{\partial x_{\rm L \nu}}
 \frac{\partial x_{\rm L \nu^{\prime}}}{\partial x_{\mu}}
 \frac{\partial \Omega^{4\tilde{\rm E}}_{{\rm L}q}(x_{\rm L})}{\partial x_{\rm L \nu^{\prime}}}
\end{eqnarray}
\begin{eqnarray}
-\frac{1}{2}m^2\sum_{p,q}\phi_p\phi_q
\Omega^{4\tilde{\rm E}}_{{\rm L}p}(x_{\rm L}) \Omega^{4\tilde{\rm E}}_{{\rm L}q}(x_{\rm L})
\},
\end{eqnarray}
\begin{eqnarray}
\label{eqn:Cvph3}
S_{3}^{(3)}=\int dx_{\rm L}^4 J_{f_{\rm L}}
\{-\frac{g_3}{3!}
\sum_{p,q,r}\phi_p\phi_q\phi_r
\Omega^{4\tilde{\rm E}}_{{\rm L}p}(x_{\rm L}) \Omega^{4\tilde{\rm E}}_{{\rm L}q}(x_{\rm L})
\Omega^{4\tilde{\rm E}}_{{\rm L}r}(x_{\rm L})
\},
\end{eqnarray}
where $J_{f_{\rm L}}$ is the Jacobian for the transformation of coordinates.
\par
The relation for rotations in Minkowski/Euclid spacetime given by
\begin{eqnarray}
 \frac{\partial x_{\rm L \nu}}{\partial x_{\mu}}
 \frac{\partial x_{\rm L \nu^{\prime}}}{\partial x_{\mu}}
=\Lambda_{\nu \mu}\Lambda_{\nu^{\prime} \mu}
=\Lambda_{\nu \mu}\Lambda_{\mu \nu^{\prime} }
=\delta_{\nu \nu^{\prime}},
\end{eqnarray}
and that for the Jacobian
\begin{eqnarray}
J_{f_{\rm L}}=1,
\end{eqnarray}
yield
\begin{eqnarray}
\label{eqn:Cv32}
\nonumber
S_{3}^{(2)}=
\int dx_{\rm L}^4
\{-\frac{1}{2}
\sum_{p,q}\phi_p\phi_q
 \frac{\partial \Omega^{4\tilde{\rm E}}_{{\rm L}p}(x_{\rm L})}{\partial x_{\rm L \nu}}
 \frac{\partial \Omega^{4\tilde{\rm E}}_{{\rm L}q}(x_{\rm L})}{\partial x_{\rm L \nu}}
\end{eqnarray}
\begin{eqnarray}
-\frac{1}{2}m^2\sum_{p,q}\phi_p\phi_q
\Omega^{4\tilde{\rm E}}_{{\rm L}p}(x_{\rm L}) \Omega^{4\tilde{\rm E}}_{{\rm L}q}(x_{\rm L})
\},
\end{eqnarray}
\begin{eqnarray}
\label{eqn:Cvph3L}
S_{3}^{(3)}=\int dx_{\rm L}^4
\{-\frac{g_3}{3!}
\sum_{p,q,r}\phi_p\phi_q\phi_r
\Omega^{4\tilde{\rm E}}_{{\rm L}p}(x_{\rm L}) \Omega^{4\tilde{\rm E}}_{{\rm L}q}(x_{\rm L})
\Omega^{4\tilde{\rm E}}_{{\rm L}r}(x_{\rm L})
\}.
\end{eqnarray}
Equations (\ref{eqn:Cv32}) and (\ref{eqn:Cvph3L}) state that the present Lagrangian formalism, wherein the fields are expressed in terms of the step-function-type basis functions in the parameter and real spacetime, is Poincar${\rm \acute{e}}$ covariant. This is due to the property of the spacetime continuum by regarding the basis function like a physical object in the spacetime continuum.
We notice that the initial and final conditions are necessarily required to be imposed at the lattice (grid) points in the parameter (real) spacetime, because the basis function such as $\Omega^{4\tilde{\rm E}}_{{\rm R}p}(x)$ is constant around the lattice point and the coefficient ($\phi_{p}$) of the basis function can be given in the region, where the basis function is constant around the lattice point.
\par

\subsection{Path integral approach to a scalar field using step-function-type basis functions \label{sec:Sec24}}

We here demonstrate the properties of the present formalism for fields that use the step-function-type basis functions with finite degrees of freedom in the spacetime continuum developed in Subsections \ref{sec:Sec21}-\ref{sec:Sec23}. An example chosen here is a self-interacting scalar field in two-dimensional (2D) Euclidean spacetime. The norm of a Euclidean vector and infinitesimal volume are
\begin{eqnarray}
x_{\mu}x_{\mu}=x_{t}x_{t}+x_{x}x_{x},
\end{eqnarray}
\begin{eqnarray}
d^2 x = dt dx,
\end{eqnarray}
and the field $\phi(x)$ means
\begin{eqnarray}
\phi(x)=\phi(t,x)=\phi(t,{\bf x}).
\end{eqnarray}
The action for $m=0$ is given by
\begin{eqnarray}
\label{eqn:Ac3}
S_{3}=S^{(2)}_{3}+S^{(3)}_{3},
\end{eqnarray}
where
\begin{eqnarray}
\label{eqn:aS32}
\nonumber
S_{3}^{(2)}
=\frac{1}{2}\int d^2x 
\partial_{\nu} \phi\partial_{\nu}\phi
\end{eqnarray}
\begin{eqnarray}
=\frac{1}{2}\int dtdx 
[
\partial_{t} \phi(x)\partial_{t} \phi(x)
+\partial_{x} \phi(x)\partial_{x} \phi(x)],
\end{eqnarray}
\begin{eqnarray}
S_{3}^{(3)}=\frac{g_{3}}{3!}\int dtdx [\phi(x)]^3.
\end{eqnarray}
The spacetime domain we are considering is $0<T_{({\rm b})0} \leq t \leq T_{({\rm b})}$, $-X_{({\rm b})} \leq x \leq X_{({\rm b})}$ in real spacetime.
\par
For the sake of simplicity, hereafter, real spacetime is treated to be identical as the parameter spacetime described in the previous section; namely, $x=x_{\rm P}$ in Eq. (\ref{eqn:DfPMP}). 
For the square lattice in the domain of the parameter spacetime, the notation of the lattice (grid) point with the index $p=(k,l)$ is
\begin{eqnarray}
x_{p}=x_{(k,l)}=(t_{k},x_{l}),
\end{eqnarray}
where $N_{t}$ and $N_{x}$ are the total numbers of lattice points along the $t$ and $x$ axes, respectively. The field is expressed in terms of step-function-type basis functions in the two-dimensional spacetime
\begin{eqnarray}
\label{eqn:S2B1}
\phi(x)
=\sum_{p} \phi_{p}\Omega^{2\tilde{\rm E}}_{p}(x)
=\sum_{k,l} \phi_{(k,l)}\Omega^{2\tilde{\rm E}}_{(k,l)}(t,x),
\end{eqnarray}
where the basis function $\Omega^{2\tilde{\rm E}}_{p}(x)$ in two-dimensional real spacetime corresponds to $\Omega^{4\tilde{\rm E}}_{p}(x_{\rm P})$ in Eq. (\ref{eqn:DeEB}) for four-dimensional parameter spacetime, and is written as
\begin{eqnarray}
\label{eqn:S2B2}
\Omega^{2\tilde{\rm E}}_{(k,l)}(t,x)=\Omega^{\tilde{\rm E}}_{k}(t)\Omega^{\tilde{\rm E}}_{l}(x),
\end{eqnarray}
where the basis function $\Omega^{\tilde{\rm E}}_{k}(t)$ in real spacetime, corresponding to the basis function $\Omega^{\tilde{\rm E}}_{k}(t_{\rm P})$ in the parameter space, is denoted by
\begin{eqnarray}
 \label{eqn:DeEBR}
\Omega^{\tilde{\rm E}}_{k}(t)=
\left\{\begin{array}{ll}
0 & \mbox{ for 
$t < t_{(k-1/2)}$  }\\
\mbox{ } & \mbox{ } \\
1 & \mbox{ for 
$t_{(k-1/2)} \leq t < t_{(k+1/2)}$ }\\
\mbox{ } & \mbox{ } \\
0 & \mbox{for 
$t_{(k+1/2)} \leq t$ }
\end{array}\right. .
\end{eqnarray}
Similarly, corresponding to the four-dimensional case in Eqs. (\ref{eqn:DB3N}) and (\ref{eqn:DB3P}) for the parameter spacetime, partial derivatives of the basis functions in real spacetime are written as (with the lattice spacing $\Delta$)
\begin{eqnarray}
\nonumber
\Omega^{\delta 1-}_{tp}(x)=
\Omega^{\delta 1-}_{t(k,l)}(t,x)
=\partial_{t}\Omega^{\tilde{\rm E}}_{k}(t)|_{t=t_{(k-1/2)}}
\Omega^{\tilde{\rm E}}_{l}(x)\Delta
\end{eqnarray}
\begin{eqnarray}
=\delta(t-t_{(k-1/2)})
\Omega^{\tilde{\rm E}}_{l}(x)\Delta,
\end{eqnarray}
\begin{eqnarray}
\nonumber
\Omega^{\delta 1+}_{tp}(x)
=-\partial_{t}\Omega^{\tilde{\rm E}}_{k}(t)|_{t=t_{(k+1/2)}}
\Omega^{\tilde{\rm E}}_{l}(x)\Delta
\end{eqnarray}
\begin{eqnarray}
=\delta(t-t_{(k+1/2)})
\Omega^{\tilde{\rm E}}_{l}(x)\Delta.
\end{eqnarray}
\par
In the kinetic terms of the action in Eq. (\ref{eqn:aS32}), 
we denote the derivative as
\begin{eqnarray}
\partial_{\nu}\sum_{p}\phi_{p}\Omega^{2\tilde{\rm E}}_{p}(x)
=\frac{\partial \sum_{p}\phi_{p}\Omega^{2\tilde{\rm E}}_{p}(x)}
{\partial x_{\nu}},
\end{eqnarray}
and then, the action for the kinetic term is expressed as
\begin{eqnarray}
\label{eqn:AcS2}
S^{(2)}_3
=\frac{1}{2}\int d^2x
[\partial_{\nu}\sum_{p}\phi_{p}\Omega^{2\tilde{\rm E}}_{p}(x)]
[\partial_{\nu}\sum_{q}\phi_{q}\Omega^{2\tilde{\rm E}}_{q}(x)].
\end{eqnarray}
Using 
\begin{eqnarray}
\partial_{\nu}\Omega^{2\tilde{\rm E}}_{p}
=\frac{1}{\Delta}\Omega^{\delta 1-}_{\nu p}-
\frac{1}{\Delta}\Omega^{\delta 1+}_{\nu p},
\end{eqnarray}
we obtain
\begin{eqnarray}
\nonumber
S^{(2)}_{3}
=\frac{1}{\Delta ^2} \{ \sum_{k,l} \sum_{K,L}
\phi_{(k,l)} \phi_{(K,L)}
\end{eqnarray}
\begin{eqnarray}
\times
[\Omega^{\delta 1-}_{\nu (k,l)}\Omega^{\delta 1-}_{\nu(K,L)}
-\Omega^{\delta 1-}_{\nu (k,l)}\Omega^{\delta 1+}_{\nu(K,L)}
-\Omega^{\delta 1+}_{\nu (k,l)}\Omega^{\delta 1-}_{\nu(K,L)}
+\Omega^{\delta 1+}_{\nu (k,l)}\Omega^{\delta 1+}_{\nu(K,L)}] \}.
\end{eqnarray}
Denoting $G_{k}(K)=G_{k,K}$, we obtain
\begin{eqnarray}
\sum_{K} \int dt G_{k}(K) \delta(t-t_{k-1/2})
\delta(t-t_{K-1/2})
=\sum_{K}\frac{1}{\Delta} G_{k,K} \delta_{k,K},
\end{eqnarray}
\begin{eqnarray}
\int dx \Omega^{\tilde{\rm E}}_{l}(x)\Omega^{\tilde{\rm E}}_{L}(x)
=\delta_{l,L}\Delta,
\end{eqnarray}
and it follows that
\begin{eqnarray}
\nonumber
S^{(2)}_{3}=
\frac{1}{2} \sum_{k,l}\sum_{K,L}
[\phi_{(k,l)} \phi_{(K,L)}
\end{eqnarray}
\begin{eqnarray}
\nonumber
\times (
2\delta_{K,k}\delta_{L,l}
-\delta_{K,k-1}\delta_{L,l}
-\delta_{K,k+1}\delta_{L,l}
\end{eqnarray}
\begin{eqnarray}
+2\delta_{K,k}\delta_{L,l}
 -\delta_{K,k}\delta_{L,l-1}
 -\delta_{K,k}\delta_{L,l+1})].
\end{eqnarray}
\par
The above quadratic part is diagonalized, using the analogy with the matrix in a vibrational problem \cite{SlaFra}. We represent $S^{(2)}_{3}$ as
\begin{eqnarray}
\label{eqn:S3Mt2}
S^{(2)}_{3}
=\sum_{p,q}\phi_{p}M^{(2)}_{pq} \phi_{q},
\end{eqnarray}
where
\begin{eqnarray}
\label{eqn:SEq01}
\nonumber
\hspace{-12ex}
M^{(2)}_{pq}
=\frac{1}{2}[
2\delta_{K,k}\delta_{L,l}
-\delta_{K,k-1}\delta_{L,l}
-\delta_{K,k+1}\delta_{L,l}
\end{eqnarray}
\begin{eqnarray}
+2\delta_{K,k}\delta_{L,l}
 -\delta_{K,k}\delta_{L,l-1}
 -\delta_{K,k}\delta_{L,l+1}].
\end{eqnarray}
The secular equation is
\begin{eqnarray}
\sum_{q}M^{(2)}_{pq}x^{(2)}_{q}
=\eta^{(2)} x^{(2)}_{p},
\end{eqnarray}
where $x^{(2)}_{q}$ and $\eta^{(2)}$ are the eigenvectors and associated eigenvalues, respectively.
We derive the $v$-th eigenvector $x^{(2)v}_{q}$ expressed by
\begin{eqnarray}
\label{eqn:eivc}
x^{(2)v}_{q}=x^{(2)v}_{(K,L)}
=\frac{1}{C^{(2)}_{\rm N}}
\sin (\vartheta_{K}j_{0}) \sin (\vartheta_{L}j_{1}),
\end{eqnarray}
with $C^{(2)}_{\rm N}$ being a normalization constant. A boundary condition such as
\begin{eqnarray}
\sin(\vartheta_{L}j_{1})=0
\hspace{3ex} \mbox{for $L=0$, $L=N_{(1)}+1$},
\end{eqnarray}
with $j_\mu$ being integers from 1 to $N_{(\mu)}$ ($N_{(\mu)}+
2
$ is the number of lattice points 
involving boundaries
in the $x_{\mu}$ axis),
achieves positive eigenvalues \cite{SlaFra}
\begin{eqnarray}
\eta^{(2)}_v=(1-c_{0})+(1-c_{1}),
\end{eqnarray}
where
\begin{eqnarray}
\label{eqn:SEq20}
c_{\mu}=\cos (z_{\mu}),
z_{\mu}=\frac{j_{\mu}\pi}{N_{(\mu)}+1}.
\end{eqnarray}
(The summation convention is not used for $(\mu)$ in parenthesis.)
The periodic boundary condition provides that
\begin{eqnarray}
c_{\mu}=\cos (z^{\prime}_{\mu}), z^{\prime}_{\mu}
=\frac{j_{\mu}(2\pi)}{N_{(\mu)}},
\end{eqnarray}
where $N_{(\mu)}$ in this case is the number of lattice points in the $x_{\mu}$ axis.
In order to avoid the appearance of a zero eigenvalue, the full non-periodic boundary condition (along all the axes) is not imposed.
\par
The self-interaction part in Eq. (\ref{eqn:Cvph3}) in terms of the basis functions is given by
\begin{eqnarray}
S_{3}
^{(3)}
\nonumber
=\frac{g_{3}}{3!}\int dtdx \sum_{p,q,r}\phi_p\phi_q\phi_r\Omega_p^{2\tilde{\rm E}}\Omega_q^{2\tilde{\rm E}}\Omega_r^{2\tilde{\rm E}}
\end{eqnarray}
\begin{eqnarray}
\nonumber
=\frac{g_{3}}{3!}\int dtdx \sum_{k,l,m,K,L,M}
\phi_{(k,K)}\phi_{(l,L)}\phi_{(m,M)}\Omega^{\tilde{\rm E}}_k (t)\Omega^{\tilde{\rm E}}_K (x)\Omega^{\tilde{\rm E}}_l (t)\Omega^{\tilde{\rm E}}_L (x)\Omega^{\tilde{\rm E}}_m (t)\Omega^{\tilde{\rm E}}_M (x)
\end{eqnarray}
\begin{eqnarray}
\label{eqn:S3Mt3}
\nonumber
=\frac{g_{3}}{3!}\sum_{k,l,m,K,L,M}
\phi_{(k,K)}\phi_{(l,L)}\phi_{(m,M)}
\delta_{k,l}\delta_{l,m}\delta_{K,L}\delta_{L,M}\Delta^2
\end{eqnarray}
\begin{eqnarray}
=\frac{g_{3}}{3!}\sum_{k,K}
(\phi_{(k,K)})^3\Delta^2
=\frac{g_{3}}{3!}\sum_{p}
(\phi_{p})^3\Delta^2.
\end{eqnarray}
\par
The transformation, corresponding to the aforementioned manipulation for the diagonalization of the matrix $M^{(2)}_{pq}$ to achieve eigenvectors described around Eq. (\ref{eqn:eivc}),
\begin{eqnarray}
\phi^{\prime}_{p}=\sum_{q}R_{pq}\phi_{q},
\end{eqnarray}
leads to
\begin{eqnarray}
\label{eqn:S22}
S_{3}^{(2)}=\sum_{p} \eta^{(2)}_{p} (\phi^{\prime}_p)^{2},
\end{eqnarray}
\begin{eqnarray}
\label{eqn:S23}
S_{3}^{(3)}=\frac{g_{3}}{3!}\sum_{q,p} (R^{-1}_{qp}\phi^{\prime}_p)^{2}\Delta^2.
\end{eqnarray}
The expectation value of the physical quantity $A_{\rm ph}$ is calculated, using the path integral with the associated weight involving the action in Eq. (\ref{eqn:Ac3}),
\begin{eqnarray}
\nonumber
<A_{\rm ph}>=\frac
{ \int D[\phi^{\prime}_{p}] (A_{\rm ph})\exp(-S_{3}) }
{ \int D[\phi^{\prime}_{p}]           \exp(-S_{3}) }
\end{eqnarray}
\begin{eqnarray}
=\frac
{
\int d\phi^{\prime}_{p_{1}}
\int d\phi^{\prime}_{p_{2}}
\int d\phi^{\prime}_{p_{3}}
\cdot\cdot\cdot\cdot
(A_{\rm ph})\exp(-S_{3})
}
{
\int d\phi^{\prime}_{p_{1}}
\int d\phi^{\prime}_{p_{2}}
\int d\phi^{\prime}_{p_{3}}
\cdot\cdot\cdot\cdot
}.
\end{eqnarray}
The above integrand is expanded in a power series
\begin{eqnarray}
\nonumber
Z_{3, {\rm N}}=\int D[\phi^{\prime}_{p}] \exp(-S_{3})
\end{eqnarray}
\begin{eqnarray}
=\int D[\phi^{\prime}_{p}] \exp(-S^{(2)}_{3})[\sum_{n}\frac{(-1)^n}{n!}(S^{(3)}_{3})^n].
\end{eqnarray}
Using Eqs. (\ref{eqn:S22})-(\ref{eqn:S23}), each one-dimensional integral of the multifold integral has the form
\begin{eqnarray}
I_{k}=\int d \phi^{\prime}_{p} (\phi^{\prime}_{p})^k
\exp[-\eta^{(2)}_p (\phi^{\prime}_{p})^{2}],
\end{eqnarray}
and amounts to
\begin{eqnarray}
I_{k}=
\left\{\begin{array}{ll}
\frac{
(2j-
1)!!
}
{2^j}
[
\frac{\pi}{(\eta^{(2)}_p)^{2j+1}}
]
^{1/2} & \mbox{ for $k=2j$}\\
\mbox{ } & \mbox{ } \\
0 & \mbox{ for $k=2j+1$}
\end{array}\right. .
\end{eqnarray}
\par
We further note that, to symmetrize the discontinuity of basis functions, we may introduce another set of basis functions, corresponding to those in Eqs. (\ref{eqn:DeEB}) and (\ref{eqn:DeEBR}), such as
\begin{eqnarray}
 \label{eqn:DeEBA}
\Omega^{\tilde{\rm E}^{\prime}}_{k}(t)=
\left\{\begin{array}{ll}
0 & \mbox{ for 
$t \leq t_{(k-1/2)}$  }\\
\mbox{ } & \mbox{ } \\
1 & \mbox{ for 
$t_{(k-1/2)} < t \leq t_{(k+1/2)}$ }\\
\mbox{ } & \mbox{ } \\
0 & \mbox{for 
$t_{(k+1/2)} < t$ }
\end{array}\right. ,
\end{eqnarray}
\begin{eqnarray}
\label{eqn:S2B2A}
\Omega^{2\tilde{\rm E}^{\prime}}_{(k,l)}(t,x)=\Omega^{\tilde{\rm E}^{\prime}}_{k}(t)\Omega^{\tilde{\rm E}^{\prime}}_{l}(x).
\end{eqnarray}
The field expansion, corresponding to that in Eq. (\ref{eqn:S2B1}), is written as
\begin{eqnarray}
\label{eqn:S2B1A}
\phi(x)
=\sum_{p} \phi_{p}^{\prime}\Omega^{2\tilde{\rm E}^{\prime}}_{p}(x)
=\sum_{k,l} \phi_{(k,l)}^{\prime}\Omega^{2\tilde{\rm E}^{\prime}}_{(k,l)}(t,x).
\end{eqnarray}
Using above relations in this note, other quantities such as the action in Eq. (\ref{eqn:AcS2}) are symmetrized, thus resulting in
\begin{eqnarray}
\label{eqn:AcS2A}
\nonumber
S^{(2)}_3=\frac{1}{2} \{ 
\frac{1}{2}\int d^2x
[\partial_{\nu}\sum_{p}\phi_{p}\Omega^{2\tilde{\rm E}}_{p}(x)]
[\partial_{\nu}\sum_{q}\phi_{q}\Omega^{2\tilde{\rm E}}_{q}(x)]
\end{eqnarray}
\begin{eqnarray}
+\frac{1}{2}\int d^2x
[\partial_{\nu}\sum_{p}\phi_{p}^{\prime}\Omega^{2\tilde{\rm E}^{\prime}}_{p}(x)]
[\partial_{\nu}\sum_{q}\phi_{q}^{\prime}\Omega^{2\tilde{\rm E}^{\prime}}_{q}(x)]
\}.
\end{eqnarray}

\subsection{Relationship between the variational approach for Hamiltonian and the path integral using step-function-type basis functions \label{sec:Sec25}}

The path integral described in the previous Subsection \ref{sec:Sec24} has a large weight at a stationary point (frequently at a local minimum such as the ground state). One of the practical approaches is the variational method for the Hamiltonian at the stationary point. Concerning the Hamiltonian formalism in field theory, Tomonaga showed that state vectors satisfy the Schr$\ddot{\rm o}$dinger equation for second quantized fields. Thus, we inspect the relationship between the variational approach for the static (time-independent) Hamiltonian and the path integral using step-function-type basis functions. The time-independent Hamiltonian of the scalar field considered here is
\begin{eqnarray}
H_{3}=H^{(2)}_{3}+H^{(3)}_{3},
\end{eqnarray}
where
\begin{eqnarray}
H_{3}^{(2)}
=\frac{1}{2}\int dx 
[\partial_{x} \phi(x)\partial_{x} \phi(x)],
\end{eqnarray}
\begin{eqnarray}
\label{eqn:SInt}
H_{3}^{(3)}=-\frac{g_{3}}{3!}\int dx [\phi(x)]^3.
\end{eqnarray}
By denoting
\begin{eqnarray}
n_{\phi}=\phi(x)\phi(x),
\end{eqnarray}
the above interaction in Eq. (\ref{eqn:SInt}) is rewritten in the form
\begin{eqnarray}
H_{3}^{(3)}=-\frac{g_{3}}{3!}\int dtdx [n_{\phi}(x)]^{3/2},
\end{eqnarray}
or it is further expanded into Taylor series
\begin{eqnarray}
H_{3}^{(3)}=-\frac{g_{3}}{3!}\int dtdx
\{
{n_{\phi}(0)}^{3/2}
+\frac{3}{2}[n_{\phi}(0)]^{1/2}n_{\phi}(x)+...
\}.
\end{eqnarray}
\par
Further, the scalar wave function is expressed using step-function-type basis functions. Corresponding to the two-dimensional case in Eq. (\ref{eqn:S2B2}), the basis functions in one-dimension are given by
\begin{eqnarray}
\label{eqn:S1B1}
\Omega^{1\tilde{\rm E}}_{(l)}(x)=\Omega^{\tilde{\rm E}}_{(l)}(x),
\end{eqnarray}
where $\Omega^{\tilde{\rm E}}_{l}(x)$ is defined in Eq. (\ref{eqn:DeEBR}). Then, the wave function has the following form
\begin{eqnarray}
\label{eqn:H1B1}
\phi(x)
=\sum_{p} \phi_{p}\Omega^{\tilde{\rm E}}_{p}(x)
=\sum_{l} \phi_{(l)}\Omega^{\tilde{\rm E}}_{(l)}(x).
\end{eqnarray}
In a similar way as Eqs.(\ref{eqn:S3Mt2}) and (\ref{eqn:S3Mt3}), it follows that
\begin{eqnarray}
\label{eqn:SHm3}
H_{3}=H^{(2)}_{3}+H^{(3)}_{3}
=\frac{1}{2}\sum_{p,q}\phi_{p}M^{\prime (2)}_{pq} \phi_{p}
-\frac{g_{3}}{3!}\sum_{p}(\phi_{p})^3\Delta,
\end{eqnarray}
where, assigning (as in higher dimensions) $p=L$; $q=l$ and $L=1, 2, ..., N_{x}$; $l=1, 2, ..., N_{x}$ ($N_{x}$ is the number of lattice (grid) points), we obtain
\begin{eqnarray}
M^{\prime (2)}_{pq}=M^{\prime (2)}_{Ll}
=\frac{1}{\Delta}
( 2\delta_{L,l} -\delta_{L,l-1} -\delta_{L,l+1} ).
\end{eqnarray}
From the normalization condition for $\phi$, the term
$\sum_{L,l}E^{\prime}(\phi_{L}\phi_{l}-1/N_{x})\delta_{L,l}\Delta
/2
$
is added to the above Hamiltonian in Eq. (\ref{eqn:SHm3}), where $E^{\prime}$ is a Lagrange multiplier.
\par
Variational calculus with respect to $\phi_{p}=\phi_{L}$ in Eq. (\ref{eqn:SHm3}) leads to the following secular equation for the $v$-th eigenvector $\phi^{v}_q=\phi^{v}_l$ associated with the $v$-th eigenenergy $E^{\prime v}$
\begin{eqnarray}
\label{eqn:SSEq01}
\sum_q M^{\prime\prime v (2)}_{pq}\phi^{v}_q=\sum_q E^{\prime v} \delta_{p,q}\phi^v_q \Delta,
\end{eqnarray}
where
\begin{eqnarray}
\label{eqn:SSMt01}
M^{\prime\prime v (2)}_{pq}=M^{\prime (2)}_{pq}
-\frac{g_{3}}{2!}
\phi^v_{q}
\delta_{p,q}\Delta.
\end{eqnarray}
The above Eq. (\ref{eqn:SSEq01}) results in
\begin{eqnarray}
\label{eqn:SSqE02}
\sum_q (\frac{1}{\Delta}M^{\prime\prime v (2)}_{pq})\phi^{v}_q=\sum_q (E^{\prime v}\delta_{p,q})\phi^v_q.
\end{eqnarray}
We note that, the above secular equation (given by Eq. (\ref{eqn:SSqE02})), for the $v$-th eigenvector $\phi^v_{q}$, involves the self-interacting $v$-th eigenvector $\phi^v_{q}$ in the potential of the matrix form $M^{\prime\prime v (2)}_{pq}$ (given by Eq. (\ref{eqn:SSMt01})). Then, the eigenvector is not directly derived from the above secular equation via the one-linear step, and self-consistent iterations are required to derive the eigenvector $\phi^v_{q}$. In other words, at the initial step, we assign $\phi^v_{p}=\phi^{v(0)}_{p}$ (with $\phi^{v(0)}_{p}$ being an initial value) for $M^{\prime\prime v (2)}_{pq}$ (given by Eq. (\ref{eqn:SSMt01}) and used in Eq. (\ref{eqn:SSqE02})); then, we derive the first-step eigenvector $\phi^{v}_q=\phi^{v(1)}_q$ from the secular equation in Eq. (\ref{eqn:SSqE02}). Similarly, at the second step, we assign $\phi^v_{p}=\phi^{v(1)}_{p}$ for $M^{\prime\prime v (2)}_{pq}$ (in Eqs. (\ref{eqn:SSMt01}) and (\ref{eqn:SSqE02})); subsequently, we derive the second-step eigenvector $\phi^{v}_q=\phi^{v(2)}_q$. These procedures are repeated until the input eigenvector in Eq. (\ref{eqn:SSMt01}) converges to coincide with the output-eigenvector solution from Eq. (\ref{eqn:SSqE02}). As evident from these manipulations, the secular equation to derive $v$-th eigenvector and the secular equation to derive $v^{\prime}$-th eigenvector differ from each other, because the self-interaction potentials $M^{\prime\prime v (2)}_{pq}$ for the $v$-th and $v^{\prime}$-th eigenvectors differ from each other. Hence, we can observe that the eigenvectors are not always orthogonal to each other, since the secular equations for these eigenvectors differ from each other.
\par
In this subsection, we finally include a numerical computational method to solve the time-dependent Schr$\ddot{\rm o}$dinger equation for imaginary time. As mentioned before, Tomonaga \cite{Tomo46} formulated that the state vector of second quantized fields satisfies the time-dependent Schr$\ddot{\rm o}$dinger equation. Generally, the Schr$\ddot{\rm o}$dinger equation is transformed to a diffusion equation, by replacing real time by imaginary time in the sense of the Wick rotation. In order to solve the diffusion equations, the author developed a method \cite{Fuku85a,Fuku85b} that obtains solutions in a short computational time with a reduced memory size. We consider the Fourier's work that a time-dependent solution is the superposition of modes with the decay constant $\lambda_{w}$. (The reduced number of the decay constants are selected as representatives in practical computations.) In the present method, we then build a set of variable time grid points $t^{({\rm D})}_{\xi}$ ($\xi=0, 1, 2, ..., N^{({\rm D})}_{t}-1$ and $t^{({\rm D})}_{0}=0$ for the number of grid points $N^{({\rm D})}_{t}$ ) to satisfy
\begin{eqnarray}
\frac{N^{({\rm D})}_{t}-\xi}{N^{({\rm D})}_{t}}=\exp(-\lambda_{w}t^{({\rm D})}_{\xi}).
\end{eqnarray}
By collecting all the time grid points, we obtain a union of all the sets with all the representatives $\lambda_{w}$. Then, calculations are performed using the above time grid points of the union.
\par
Concerning spatial grid points \cite{Fuku85a,Fuku85b}, an optional set of the points $x_{\eta}$ ($\eta=0, 1, 2, ..., N^{({\rm D})}_{x}$) near a boundary at the coordinate origin along such as the x-axis is chosen to satisfy
\begin{eqnarray}
x_{\eta}=L_{x}
\left (
\frac{\eta}{N^{({\rm D})}_{x}}
\right )
^{\beta},
\end{eqnarray}
where $\beta$ is such as around 1/2, and $L_{x}$ is the length of the considering region. Thus, a fast computation with a reduced memory size is attainable using the aforementioned time and space grid points.

\section{General properties of gauge fields \label{sec:Sec3}}

\subsection{
Abelian gauge 
fields \label{sec:Sec31}}

An electron in the scheme of quantum electrodynamics (QED) has an intrinsic magnetic moment of up or down spin, denoted by $\uparrow$ or $\downarrow$, as the internal degrees of freedom. The wave function $\psi_{\rm e}(x)$ of the electron with these degrees  is described in terms of the components $\psi_{\rm e \uparrow}(x)$ and $\psi_{\rm e \downarrow}(x)$ of up and down spins
\begin{eqnarray}
\psi_{\rm e}(x)=
\left(\begin{array}{c}
\psi_{\rm e \uparrow}(x) \\
\psi_{\rm e \downarrow}(x) \\
\end{array}\right).
\end{eqnarray}
Besides the intrinsic magnetic moment, the electron has internal degrees of freedom to be a particle or antiparticle, and the above two-component wave function is generalized to spinor. The degrees of freedom for the spin and particle/antiparticle states are unified into a single spinor to completely describe the relativistic electron. In contrast, different kinds of degrees of freedom, such as color for strong interaction between fundamental fermion particles in the scheme of quantum chromodynamics (QCD), are represented by a vector whose components are different kinds of spinors. The rotational symmetry operations on vectors composed of single-spinor and multi-spinors are represented by the matrices ${\rm U(1)}$ and ${\rm SU}(N)$ ($N=2,3,...$), respectively. QED for a single elementary fermion (electron) is represented by the vector composed of a single spinor associated with the ${\rm U(1)}$ rotational matrix. Meanwhile, as mentioned later, QCD, for such as three elementary fermions, is represented by the vector composed of such as three different kinds of spinors associated with the matrix ${\rm SU(3)}$.
\par
The spinor of the electron in QED has an ambiguity, allowing the local gauge transformation. Consistent with the aforementioned symmetry matrix for the vector composed of spinor, the gauge transformation contains the matrix identical to the symmetry matrix operating on the vector. In QED, the symmetry operation matrix is ${\rm U}(1)$ for the electron wave function (with spin and particle/antiparticle states). The gauge transformation for the electron spinor $\psi_{\rm e}^{(0)}(x)$ (superscript (0) distinguishes Abelian QED quantities from non-Abelian quantities) is then written as
\begin{eqnarray}
\psi_{\rm e}^{(0)}(x) \rightarrow \psi_{\rm e}^{(0)}(x)\exp
(i
\alpha^{(0)}(x)U^{(1)}),
\end{eqnarray}
where $\alpha^{(0)}(x)$ is a gauge parameter and $U^{(1)}$ is a ${\rm U}(1)$ matrix. The gauge-covariant derivative, for the gauge-invariant formalism, is defined by
\begin{eqnarray}
D^{(0)}_{\mu}(x)=I^{(1)}\partial_{\mu}
-
ieA^{(0)}_{\mu}(x)U^{(1)},
\end{eqnarray}
where $I^{(1)}$ is a unit matrix in the ${\rm U}(1)$ case, $e$ is coupling constant and $A^{(0)}_{\mu}$ is the ${\rm U}(1)$ gauge field introduced to hold the gauge invariance. The field tensor is given by
\begin{eqnarray}
\nonumber
F^{(0)}_{\mu \nu}=\frac{
i
}{e}[D^{(0)}_{\mu}(x),D^{(0)}_{\nu}(x)]
\end{eqnarray}
\begin{eqnarray}
\nonumber
=\frac{
i
}{e}(D^{(0)}_{\mu}(x)D^{(0)}_{\nu}(x)-D^{(0)}_{\nu}(x)D^{(0)}_{\mu}(x))
\end{eqnarray}
\begin{eqnarray}
=\partial_{\mu}A^{(0)}_{\nu}(x)-\partial_{\nu}A^{(0)}_{\mu}(x).
\end{eqnarray}
\par
Using the property of trace for the unitary (gauge) transformation of the matrix $A_{\rm ph,M}(x)$, which describes a physical quantity,
\begin{eqnarray}
{\rm Tr}[(U^{(1)})^{-1}A_{\rm ph,M}(x)U^{(1)}]={\rm Tr}(A_{\rm ph,M}(x)),
\end{eqnarray}
the trace of the field tensor satisfies the following gauge invariance
\begin{eqnarray}
{\rm Tr}(F^{(0)}_{\mu \nu})={\rm Tr}[(U^{(1)})^{-1}F^{(0)}_{\mu \nu}U^{(1)}].
\end{eqnarray}
For the infinitesimal gauge transformation of the gauge field, we have
\begin{eqnarray}
A^{(0)}_{\mu}(x) \rightarrow A^{(0)}_{\mu}(x) + (1/e)\partial_{\mu} \alpha^{(0)}(x).
\end{eqnarray}
Hence, the following Lagrangian is invariant under the gauge transformation
\begin{eqnarray}
{\cal L}^{(0)}=\psi^{(0)\dagger}(x
)i\gamma
_{\mu}D^{(0)}_{\mu}(x)\psi^{(0)}
(x)
+\frac{1}{4}{\rm Tr}(F^{(0)}_{\mu \nu}F^{(0)}_{\mu \nu}),
\end{eqnarray}
where $\psi^{(0)\dagger}(x)$ is the adjoint of $\psi^{(0)}(x)$ and $\gamma_{\mu}$ are the Dirac matrices.
\par

\subsection{Non-Abelian gauge fields \label{sec:Sec32}}

The non-Abelian gauge field $A_{\mu}(x)$ (in Euclidean spacetime with the norm $x_{\mu}x_{\mu}=x_{0}x_{0}+x_{1}x_{1}+x_{2}x_{2}+x_{3}x_{3}$ at time $x_{0}=t$) has the following form
\begin{eqnarray}
\label{eqn:NAfD}
A_{\mu}(x)=A_{\mu}(t,x,y,z)
=A_{\mu}(t,{\bf x})=
\sum_{a}A_{\mu}^{a}(x)T^{a},
\end{eqnarray}
where matrices $T^{a}$ generate the following Lie algebra
\begin{eqnarray}
\label{eqn:LiabD}
[T^{a},T^{b}]=\sum_{c}if^{abc}T^{c}.
\end{eqnarray}
Here, $f^{abc}$ are the structure constants, and the Lie group indices such as $a$, $b$ and $c$ run from 1 to $N$ for the ${\rm SU}(N)$ gauge field, whose dimension is $N_{\rm D}=N^2-1$.
The matrices $T^{\prime a}=2T^{a}$ for SU(2) are represented as
\par
\begin{eqnarray}
\begin{minipage}{3cm}
\begin{eqnarray}
\label{eqn:LMT2}
\nonumber
T^{\prime 1}=
\left[\begin{array}{cc}
0 & 1 \\
1 & 0 \\
\end{array}\right],
\end{eqnarray}
\end{minipage}
\begin{minipage}{3cm}
\begin{eqnarray}
\nonumber
T^{\prime 2}=
\left[\begin{array}{cc}
0 & -i \\
i &  0 \\
\end{array}\right],
\end{eqnarray}
\end{minipage}
\begin{minipage}{3cm}
\begin{eqnarray}
\nonumber
T^{\prime 3}=
\left[\begin{array}{cc}
1 &  0 \\
0 & -1 \\
\end{array}\right],
\end{eqnarray}
\end{minipage}
\end{eqnarray}
\begin{eqnarray}
\nonumber
\end{eqnarray}
{\flushleft with the following totally antisymmetric real structure constants}
\begin{eqnarray}
\label{eqn:SCN2}
f_{123}=1.
\end{eqnarray}
\par
In the SU(3) case, the Lie matrices are
\begin{eqnarray}
\begin{minipage}{3.5cm}
\begin{eqnarray}
\nonumber
\label{eqn:LMT3}
T^{\prime 1}=
\left[\begin{array}{ccc}
0 & 1 & 0 \\
1 & 0 & 0 \\
0 & 0 & 0 \\
\end{array}\right],
\end{eqnarray}
\end{minipage}
\begin{minipage}{3.5cm}
\begin{eqnarray}
\nonumber
T^{\prime 2}=
\left[\begin{array}{ccc}
0 & -i & 0 \\
i &  0 & 0 \\
0 &  0 & 0 \\
\end{array}\right],
\end{eqnarray}
\end{minipage}
\begin{minipage}{3.5cm}
\begin{eqnarray}
\nonumber
T^{\prime 3}=
\left[\begin{array}{ccc}
1 &  0 & 0 \\
0 & -1 & 0 \\
0 &  0 & 0 \\
\end{array}\right],
\end{eqnarray}
\end{minipage}
\nonumber
\end{eqnarray}
\begin{eqnarray}
\begin{minipage}{3.5cm}
\begin{eqnarray}
\nonumber
T^{\prime 4}=
\left[\begin{array}{ccc}
0 & 0 & 1 \\
0 & 0 & 0 \\
1 & 0 & 0 \\
\end{array}\right],
\end{eqnarray}
\end{minipage}
\begin{minipage}{3.5cm}
\begin{eqnarray}
\nonumber
T^{\prime 5}=
\left[\begin{array}{ccc}
0 & 0 & -i \\
0 & 0 &  0 \\
i & 0 &  0 \\
\end{array}\right],
\end{eqnarray}
\end{minipage}
\begin{minipage}{3.5cm}
\begin{eqnarray}
\nonumber
T^{\prime 6}=
\left[\begin{array}{ccc}
0 & 0 & 0 \\
0 & 0 & 1 \\
0 & 1 & 0 \\
\end{array}\right],
\end{eqnarray}
\end{minipage}
\nonumber
\end{eqnarray}
\begin{eqnarray}
\begin{minipage}{3.5cm}
\begin{eqnarray}
\nonumber
T^{\prime 7}=
\left[\begin{array}{ccc}
0 & 0 &  0 \\
0 & 0 & -i \\
0 & i &  0 \\
\end{array}\right],
\end{eqnarray}
\end{minipage}
\begin{minipage}{3.5cm}
\begin{eqnarray}
\nonumber
T^{\prime 8}=\frac{1}{\sqrt{3}}
\left[\begin{array}{ccc}
1 & 0 &  0 \\
0 & 1 &  0 \\
0 & 0 & -2 \\
\end{array}\right],
\end{eqnarray}
\end{minipage}
\end{eqnarray}
\begin{eqnarray}
\nonumber
\end{eqnarray}
with the following totally antisymmetric real structure constants
\begin{eqnarray}
\label{eqn:SCN3}
\nonumber
f_{123}=1,
f_{147}= \frac{1}{2},
f_{156}=-\frac{1}{2},
f_{246}= \frac{1}{2},
f_{257}= \frac{1}{2},
f_{345}= \frac{1}{2},
f_{367}=-\frac{1}{2},
\end{eqnarray}
\begin{eqnarray}
\nonumber
f_{458}=-\frac{\sqrt{3}}{2}, 
f_{478}=-\frac{\sqrt{3}}{2},
\end{eqnarray}
\begin{eqnarray}
f^{abc}=0 \hspace{4ex}\mbox{for others}.
\end{eqnarray}
\par
As mentioned in the previous subsection, the fundamental fermion particles in the non-Abelian system have color degrees of freedom, and the wave function with these degrees of freedom has $N$ components of spinors. The system has SU($N$) symmetry, and the gauge transformation under this symmetry extends U(1) matrix with a scalar gauge parameter to  the SU($N$) Lie matrices with multi-gauge parameters for the gauge transformation to obtain
\begin{eqnarray}
\psi(x) \rightarrow \psi(x)\exp
(i
\sum_{a}\alpha^{a}(x)T^{a}).
\end{eqnarray}
Here, $\alpha^{a}(x)$ are gauge parameters and the gauge covariant derivative is of the form
\begin{eqnarray}
D_{\mu}(x)=I\partial_{\mu}+ig\sum_{a}A^{a}_{\mu}(x)T^{a},
\end{eqnarray}
where $I$ is a unit matrix and $g$ is the coupling constant in the non-Abelian case. The field tensor of the non-Abelian field is described as
\begin{eqnarray}
F_{\mu \nu}=\frac{-i}{g}[D_{\mu}(x),D_{\nu}(x)],
\end{eqnarray}
with its component
\begin{eqnarray}
F_{\mu \nu}^a=\partial_{\mu}A^a_{\nu}(x)-\partial_{\nu}A^a_{\mu}(x)
-
g\sum_{b,c} f^{abc} 
A^{b}_{\mu}A^{c}_{\nu}.
\end{eqnarray}
The infinitesimal gauge transformation of the gauge field is generalized to
\begin{eqnarray}
\delta A^{a}_{\mu}(x)=-(1/g)\partial_{\mu} \delta\alpha^{a}(x)
-\sum_{b,c} f^{bca}\delta\alpha^{b}(x)A^{c}_{\mu}(x).
\end{eqnarray}
\par
Thus, the non-Abelian Lagrangian is written in the form
\begin{eqnarray}
{\cal L}_{\rm YM}=\psi^{\dagger}(x
)i\gamma
_{\mu}D_{\mu}(x)\psi
(x)
+\frac{2}{4}{\rm Tr}(F_{\mu \nu}F_{\mu \nu}),
\end{eqnarray}
which is invariant under the gauge transformation.
In the Feynman gauge
\begin{eqnarray}
{\cal L_{\rm F}}=\frac{1}{2}\sum_{a}(\partial_{\mu} A^{a}_{\mu})^{2},
\end{eqnarray}
the action of the gauge field is expressed as
\begin{eqnarray}
\nonumber
S=\int d^4x {\cal L}=\frac{1}{2}\sum_{a} \int d^4x 
(\partial_{\nu} A^{a}_{\mu}\partial_{\nu}A^{a}_{\mu})
\end{eqnarray}
\begin{eqnarray}
\nonumber
-\frac{1}{2} \sum_{a} \int d^4x 
(\partial_{\nu}(A^{a}_{\mu}\partial_{\mu}A^{a}_{\nu}
- A_{\nu}^{a}\partial_{\mu}A^{a}_{\mu} ))
-g \sum_{a,b,c} f^{abc} \int d^4x
(A^{b}_{\mu}A^{c}_{\nu}\partial_{\mu}A^{a}_{\nu})
\end{eqnarray}
\begin{eqnarray}
+\frac{g^{2}}{4} \sum_{a,b,c,d,e} f^{abc}f^{ade} \int d^4x 
(A^{b}_{\mu}A^{c}_{\nu}A^{d}_{\mu}A^{e}_{\nu}).
\end{eqnarray}
The above action consists of the kinetic term and the self-interacting cubic and quartic terms. The second term in the Feynman gauge is dropped owing to the vanishing surface integration.
\par

\section{Fermion confinement by non-Abelian gauge field \label{sec:Sec4}}

\subsection{Classical solution of non-Abelian gauge field as a vacuum \label{sec:Sec41}}

Before describing the present new solution of the classical non-Abelian Yang-Mills equation, which can be regarded as a vacuum, we briefly summarize the solving procedure.
The process is purely mathematical, such that the physical meaning of the mathematical object is considered after the manipulation. (The physical properties of the present confinement mechanism may differ from those of the dual superconductor model (by other researchers), for which the deconfinement occurs above the superconducting critical temperature and gauge fields with the energy less than superconducting gap pass through superconductors.)
Our solution of the non-Abelian gauge field is composed of a classical field and quantum fluctuations. The classical field is described by such as the Lie group involving a SU(2) subgroup with finite field-amplitudes embedded in the concerning group (such as SU($N$)) with zero field-amplitudes, but the quantum fluctuations are described by the concerning group (such as SU($N$)) without a restriction unlike the former classical fields.
We denote classical fields as $A_{(\rm C) \mu}$; then, using Eq. (\ref{eqn:NAfD}), the components of the present solution of the SU(2) field embedded in the SU($N$) field are written in the form
\begin{eqnarray}
\label{eqn:Csol}
A_{(\rm C) \mu}(x)=
\left\{\begin{array}{ll}
A^{a}_{(\rm C) \mu}(x)T^{a}=\lambda^{a}\tilde{A}_{(\rm C) \mu}(x)T^{a}
 & \mbox{ for } a=1, 2, 3 \\
\mbox{ } & \mbox{ } \\
0 &  \mbox{ for } a>3
\end{array}\right. ,
\end{eqnarray}
where $\lambda^{a}$ are real constants.
Then, self-interaction terms of the non-Abelian Yang-Mills equation vanish owing to the antisymmetry of Lie algebraic structure constants. We further expect that the classical Wilson loop, which is the trace of an exponential function of line integrals along the time axis (line integrals along the spatial axis are set to zero), has the following form (involving a constant
$a
_{\rm c}
$
mentioned later)
\begin{eqnarray}
\label{eqn:ecWL}
\nonumber
W_{\rm C} \approx \cos \{ \arccos [\exp(-a
_{\rm c}
xt)] \}
\end{eqnarray}
\begin{eqnarray}
=\cos \{ \cos^{-1} [\exp(-a
_{\rm c}
xt)] \}
=\exp(-a
_{\rm c}
xt).
\end{eqnarray}
Then, the above Wilson loop exhibits the area law
($a
_{\rm c}
xt$ is proportional to the area $xt$)
indicating the linear potential between the fundamental particle-antiparticle fermions.
\par
The relation in Eq. (\ref{eqn:ecWL}) is the essence of the confinement, and the present solution must reproduce this relation. Since the above vector potential (localized function in the present solution explained later) in Eq. (\ref{eqn:Csol}) do not satisfy the non-Abelian Yang-Mills equation, another vector potential (unlocalized functions in the present solution mentioned later) is added to complete the solution. Since the former vector potential yields the area law, the source of the latter vector potential is set to make no contribution to the Wilson loop.
Usually, the quantum Abelian gauge field is formulated around the classical zero field in the path integral. In contrast, the non-Abelian gauge field can be constructed around a finite non-pertubative classical field as a vacuum to decrease the system energy, along with quantum fluctuations. The quantum field (fluctuation expressed in terms of the step-function-type basis functions given in Section \ref{sec:Sec2}) has perturbative properties for a small lattice spacing.
\par
Now, let us consider a fermion-antifermion pair created in the form of a wave
packet involving soliton-like objects at the origin of spacetime coordinates during a quantum process in a vacuum described by a classical field. We note that the non-Abelian phenomena have the scale-invariance, and the scale-invariant energy is not predicted by the theory but is a given condition. In the center-of-mass frame, the fermion and antifermion are located at opposite positions with respect to the origin in $x$-axis ($y=z=0$). The Euclidean classical field configuration consists of the localized function $A^{a}_{{\rm (CL)}\mu}(x)$ and the unlocalized function $A^{a}_{{\rm (CU)}\mu}(x)$
\begin{eqnarray}
\label{eqn:CSo01}
A^{a}_{\rm (C)\mu}(x)=A^{a}_{{\rm (CL)}\mu}(x)+A^{a}_{{\rm (CU)}\mu}(x).
\end{eqnarray}
The classical localized function (object of solitonlike wave-packet) is given in the region, denoted as $0 < T_{\rm (b
)0
}\leq \epsilon_{t}\leq t \leq T_{\rm (b)}$, $|x| \leq X_{\rm (b)}$, $|y| \leq Y_{\rm (b)}$ and $|z| \leq Z_{\rm (b)}$, where the subscript $\rm (b)$ indicates boundaries and $\epsilon_{t}$ is the scale-invariant time to be described in detail later in the paper.
The time component of the field $A^{a}_{{\rm (CL)}t}$ has the following form
\begin{eqnarray}
\label{eqn:CSUL1l}
A^{a}_{{\rm (CL)}t}(t,{\bf x})=
\lambda^{a}P_{(0)}(t,x)w_{t}(y)w_{t}(z) \mbox{ for $x \geq \epsilon_{x}$},
\end{eqnarray}
where $\epsilon_{x}$ is an infinitesimal positive quantity (putting $\epsilon_{x} \rightarrow 0$ after the calculations).
(A more general rotationally covariant form, which is obtained by the Lorentz transformation, in Minkowski/Euclid spacetime is found in the previous paper \cite{Fuku14}.)
In Eq. (\ref{eqn:CSUL1l}) (which we are considering), the function $P_{(0)}(t,x)$ is written as
\begin{eqnarray}
P_{(0)}(t,x)=\frac{1}{2}k^{-1/2}_{N_{\rm D}}h(t,x),
\end{eqnarray}
where
\begin{eqnarray}
\label{eqn:hDef}
h(t,x)=  \frac{ - a_{\rm c} x \exp(-a_{\rm c}|tx|) }
{ [1-\exp(-2a_{\rm c}|tx|)]^{1/2} }
\hspace{2ex}
\mbox{ for $x \geq \epsilon_{x} $},
\end{eqnarray}
which is antisymmetric as
\begin{eqnarray}
\label{eqn:hant}
h(t,-x)=-h(t,x)
\hspace{2ex}
\mbox{ for $x < -\epsilon_{x} $}.
\end{eqnarray}
(The derivative is defined in the region where the function is given.) The coefficient $k_{N_{\rm D}}$ and the parameter $a_{c}$ are determined later. 
The function such as $w_{t}(y)$ is defined as
\begin{eqnarray}
\label{eqn:wyzD}
w_{t}(y)=w(y)=
\left\{\begin{array}{ll}
+\exp(-\frac{y}{d}) & \mbox{ for $y \geq \epsilon_{y}$} \\
 &  \\
-\exp(+\frac{y}{d}) & \mbox{ for $y < -\epsilon_{y}$} \\
\end{array}\right. ,
\end{eqnarray}
where $\epsilon_{y}$ is an infinitesimal positive quantity (putting $\epsilon_{y} \rightarrow 0$ after calculations). The symbol $d$ denotes the thin thickness of the $xt$ sheet, and taken to be $d \rightarrow 0$ after calculations to reduce the energy of the classical localized function in Eq. (\ref{eqn:CSo01}).
\par
The other components of the classical localized function are given by
\par
\begin{eqnarray}
\label{eqn:CLFx}
A^{a}_{{\rm (CL)}x}(t,{\bf x})=0,
\end{eqnarray}
\begin{eqnarray}
\label{eqn:CLFy}
A^{a}_{{\rm (CL)}y}(t,{\bf x})=
\lambda^a (\frac{d}{2})(\partial_{t}P_{(0)})w_{y}(y)w(z),
\end{eqnarray}
\begin{eqnarray}
\label{eqn:CLFz}
A^{a}_{{\rm (CL)}z}(t,{\bf x})=
\lambda^a (\frac{d}{2})(\partial_{t}P_{(0)})w(y)w_{z}(z),
\end{eqnarray}
where
\begin{eqnarray}
w_{y}(y)=
\left\{\begin{array}{ll}
+w(y) & \mbox{ for $y \geq \epsilon_{y}$} \\
 &  \\
-w(y) & \mbox{ for $y < -\epsilon_{y}$} \\
\end{array}\right. ,
\end{eqnarray}
\begin{eqnarray}
w_{z}(z)=
\left\{\begin{array}{ll}
+w(z) & \mbox{ for $z \geq \epsilon_{z}$} \\
 &  \\
-w(z) & \mbox{ for $z < -\epsilon_{z}$} \\
\end{array}\right. ,
\end{eqnarray}
($\epsilon_{z}$ is an infinitesimal positive quantity, putting $\epsilon_{
z
} \rightarrow 0$ after calculations).
These classical localized functions (in such as Eq. (\ref{eqn:CSo01})) satisfy the following Lorentz condition.
\begin{eqnarray}
\partial_{\mu}A^{a}_{{\rm (CL)}\mu}(x)=0.
\end{eqnarray}
\par
The source charge density of the classical unlocalized function $A^{a}_{{\rm (CU)}\mu}(x)$ in Eq. (\ref{eqn:CSo01}) for $t \geq T_{(\rm b
)0
}$ is given by
\begin{eqnarray}
\rho^{a}_{t}(t,x,y,z)=
\left\{\begin{array}{l}
\lambda^a [
(\frac{1}{2}k_{N}^{-1/2})Q(t,x)
-\frac{2}{d^2}P_{(0)}(t,x)]w(y)w(z)\\
\mbox{ }\\
\hspace{26.5ex}
\mbox{ for $\epsilon_{x} \leq x \leq X_{c}$}\\
\mbox{ }\\
0
\hspace{25.5ex}
\mbox{ for $x > X_{c}$}\\
\end{array}\right. ,
\end{eqnarray}
where $X_{c}$ is a charge (wave packet) size, and
\begin{eqnarray}
Q(t,x)=a_{\rm c}^{3}x^{3}Q_{1}(t,x) -2a_{\rm c}^{2}tQ_{2}(t,x)
+a_{\rm c}^{3}t^{2}xQ_{1}(t,x),
\end{eqnarray}
with
\begin{eqnarray}
Q_{1}(t,x)=Q_{1/2}(t,x)+4Q_{3/2}(t,x)
+3Q_{5/2}(t,x), 
\end{eqnarray}
\begin{eqnarray}
Q_{2}(t,x)=Q_{1/2}(t,x)+Q_{3/2}(t,x), 
\end{eqnarray}
\begin{eqnarray}
Q_{1/2}(t,x)=\frac{\exp(-a_{\rm c}|tx|)}{[1-\exp(-2a_{\rm c}|tx|)]^{1/2}},
\end{eqnarray}
\begin{eqnarray}
Q_{3/2}(t,x)=\frac{\exp(-3a_{\rm c}|tx|)}{[1-\exp(-2a_{\rm c}|tx|)]^{3/2}},
\end{eqnarray}
\begin{eqnarray}
Q_{5/2}(t,x)=\frac{\exp(-5a_{\rm c}|tx|)}{[1-\exp(-2a_{\rm c}|tx|)]^{5/2}}.
\end{eqnarray}
The other charge components are in the forms of
\begin{eqnarray}
\label{eqn:Rhox}
\rho^{a}_{x}(t,{\bf x})=0,
\end{eqnarray}
\begin{eqnarray}
\rho^{a}_{y}(t,{\bf x})=-\frac{2}{d^2}P_{(0)}(t,x
)w
(y)w(z),
\end{eqnarray}
\begin{eqnarray}
\rho^{a}_{z}(t,{\bf x})=-\frac{2}{d^2}P_{(0)}(t,x
)w
(y)w(z).
\end{eqnarray}
\par
In contrast, the unlocalized function in Eq. (\ref{eqn:CSo01}) is given by
\begin{eqnarray}
\label{eqn:CUF1}
\nonumber
A^{a}_{{\rm (CU)}\mu}(t,{\bf x})=
\int_{
T_{\rm (b
)0
}
}^{t} 
dt_{s} \int_{-X_{\rm (b)}}^{X_{\rm (b)}} dx_{s}
\int_{-Y_{\rm (b)}}^{Y_{\rm (b)}} dy_{s} \int_{-Z_{\rm (b)}}^{Z_{\rm (b)}} dz_{s}
\end{eqnarray}
\begin{eqnarray}
\times
G_{4}(t,{\bf x};t_{s},{\bf x}_{s})\rho^{a}_{\mu}(t_{s},{\bf x}_{s}),
\end{eqnarray}
using the Green's function
\begin{eqnarray}
\label{eqn:CSo21}
G_{4}(t,{\bf x};t_{s},{\bf x}_{s})=\frac{1}{4\pi^{2}}
\frac{1}{(t-t_{s})^2+(x-x_{s})^2+(y-y_{s})^2+(z-z_{s})^2}.
\end{eqnarray}
In the denominator of the above integrand, $(t-t_{s})^2 + (x-x_{s})^2 +(y-y_{s})^2 +(z-z_{s})^2$ is replaced by $(t-t_{s})^2 + (x-x_{s})^2 +(y-y_{s})^2 +(z-z_{s})^2 +\epsilon_{\rm Gr}$, where $\epsilon_{\rm Gr}$ is an infinitesimal positive quantity (putting $\epsilon_{\rm Gr} \rightarrow 0$ after calculations).
This antisymmetric relation
\begin{eqnarray}
\rho^{a}_{t}(t_{s},x_{s},-y_{s},z_{s})=-\rho^{a}_{t}(t_{s},x_{s},y_{s},z_{s}),
\end{eqnarray}
\begin{eqnarray}
\rho^{a}_{t}(t_{s},x_{s},y_{s},-z_{s})=-\rho^{a}_{t}(t_{s},x_{s},y_{s},z_{s}),
\end{eqnarray}
leads to the cancellation of charge contributions to the Green's function integral at y = 0 (z = 0), and we derive
\begin{eqnarray}
\label{eqn:CUF2}
A^{a}_{{\rm (CU)}t}(t,{\bf x})=0
\hspace{2ex}
\mbox{ at $y=0$ $(z=0)$}.
\end{eqnarray}
It is found that the classical localized function satisfies $\partial^{2}_{\mu}A^{a}_{{\rm (CL)}\nu}(t,{\bf x})=-\rho^{a}_{\nu}(t,{\bf x})$.
\par
The classical non-linear Yang-Mills equation for non-Abelian gauge fields has the form
\begin{eqnarray}
\partial_{\mu}F_{\rm (C)\mu\nu}+ig_{\rm c}
[A_{\rm (C)\mu},F_{\rm (C)\mu\nu}]=0,
\end{eqnarray}
with $g_{\rm c}$ being the classical coupling constant.
The restricted form of the vector potential in
Eq. (\ref{eqn:CSUL1l}), Eq. (\ref{eqn:CLFx}), Eq. (\ref{eqn:CLFy}) and Eq. (\ref{eqn:CLFz}) satisfies
\begin{eqnarray}
A^{b}_{\rm (C)\mu}A^{c}_{\rm (C)\nu}
=\lambda^b \lambda^c \tilde{A}_{\rm (C)\mu} \tilde{A}_{\rm (C)\nu}
=A^{c}_{\rm (C)\nu}A^{b}_{\rm (C)\mu}.
\end{eqnarray}
The above equality and aforementioned antisymmetry of the structure constant cause the vanishing of the self-interacting terms $-g_{\rm c } \sum_{a,b,c} f^{abc} (A^{b}_{\mu}A^{c}_{\nu}\partial_{\mu}A^{a}_{\nu})$ and $(g_{\rm c}^{2}/4) [\sum_{a,b,c,d,e} f^{abc}f^{ade}$ $(A^{b}_{\mu}A^{c}_{\nu}A^{d}_{\mu}A^{e}_{\nu})]$ in $F^a_{\mu\nu}$. Owing to the following cancelation of the charges of the classical localized and unlocalized functions,
\begin{eqnarray}
\label{eqn:CFEQ1}
\partial^{2}_{\mu}A^{a}_{{\rm (CL)}\nu}(t,{\bf x})
=-\rho^{a}_{\nu}(t,{\bf x}),
\end{eqnarray}
\begin{eqnarray}
\label{eqn:CFEQ2}
\partial^{2}_{\mu}A^{a}_{{\rm (CU)}\nu}(t,{\bf x})
=\rho^{a}_{\nu}(t,{\bf x}),
\end{eqnarray}
the total classical vector potential in Eq. (\ref{eqn:CSo01}) satisfies the simplified linear equation
\begin{eqnarray}
\partial^{2}_{\mu}A^{a}_{\rm (C)\nu}=0.
\end{eqnarray}
\par 

\subsection{Classical Wilson loop \label{sec:Sec42}}

It is known that the static Coulomb interaction potential (in quantum electrodynamics) between charged particles in the Coulomb gauge is canceled by the retarded potential and the dynamical Coulomb potential then appears. This implies that the interaction potential contains the dynamical exchange of interaction quanta during a long time period. 
We can then expect that the dynamical interaction (in the non-Abelian case) between a charged particle and charged antiparticle is evaluated by a forward integral of the vector potential parallel to the time-axis (during a long time period) for the particle, and the corresponding backward integral for the antiparticle.
For this purpose, the Wilson loop
\begin{eqnarray}
W_{\rm L}={\rm Tr}[\exp (-ig \oint dx_{\mu}A_{\mu}(x) )],
\end{eqnarray}
along a closed path of a particle and an antiparticle is suitable.
The operator
\begin{eqnarray}
\nonumber
P_{\rm W}(x,dx)=\exp (-ig dx_{\mu} A_{\mu}(x) )
\end{eqnarray}
\begin{eqnarray}
=\exp (-ig dx_{\mu} \sum_{a} A^a_{\mu}(x)T^a ),
\end{eqnarray}
on the curve in spacetime changes the wave function as
\begin{eqnarray}
P_{\rm W}(x,dx)\psi(x)=\psi(x+dx).
\end{eqnarray}
This operator $P_{\rm W}(x,dx)$ is the integral form of the covariant derivative $D_{\mu}=I\partial_{\mu}+ig\sum_{a}A^a_{\mu}T^a$, where the wave function corresponds to a vector in curved space and the vector potential is regarded as a quantity, which causes a change in the vector due to the curvature. The change in the wave function by the operator $P_{\rm W}(x,dx)$ contains SU($N$) matrices, corresponding to a rotation matrix for a vector in curved space. Owing to the trace property of the physical quantity $A_{\rm ph}(x)$ (as a function of a spacetime point) for the gauge transformation by a unitary matrix $U$, denoted as
\begin{eqnarray}
{\rm Tr}(U^{-1}A_{\rm ph}(x)U)={\rm Tr}(A_{\rm ph}(x)),
\end{eqnarray}
the trace of the Wilson (closed) loop, whose start point is identical to the end point, is gauge invariant and extracts the interaction potential. This loop integral contains the large contributions from long time integrals of the vector potentials parallel to the time-axis at the locations of the particle and antiparticle. We note that the vector potential in the Wilson loop is the sum of the dynamical fields created by a gauge source and antisource. These two fields that are created are generally not superposed, because of the nonlinearity of the non-Abelian Yang-Mills equation. The specific linearity in the present classical-field case mentioned in the previous Subsection \ref{sec:Sec41}, which this article concerns, makes it possible to analytically derive the Wilson loop.
\par
We write the Wilson loop of the classical field $A_{({\rm C})\mu}(x)$ in the form
\begin{eqnarray}
\label{eqn:CWLP}
W_{\rm (C)}={\rm Tr}[\exp (-ig \oint dx_{\mu}A_{({\rm C})\mu}(x) )].
\end{eqnarray}
The Wilson loop we consider here is a line integral along the closed loop of a rectangle whose sides are parallel to the $x$- or $t$-axis. From Eq. (\ref{eqn:CLFx}), the line integrals along the $x$-axis yield
\begin{eqnarray}
\label{eqn:WILX}
I^{a(1)}_{{\rm (CL)}}=
\int_{x_{1}}^{x_{2}} dx A^{a}_{{\rm (CL)}x} |_{t=t_{1}, y=0, z=0}=0,
\end{eqnarray}
\begin{eqnarray}
I^{a(3)}_{{\rm (CL)}}=
\int_{x_{2}}^{x_{1}} dx A^{a}_{{\rm (CL)}x} |_{t=t_{2}, y=0, z=0} = 0,
\end{eqnarray}
whereas, using Eqs. (\ref{eqn:CSUL1l})-(\ref{eqn:wyzD}), the integrals along the $t$-axis become
\begin{eqnarray}
\nonumber
I^{a(2)}_{{\rm (CL)}}=\int_{t_{1}}^{t_{2}} dt
A^{a}_{{\rm (CL)}t} |_{x=x_{2}, y=0, z=0}
\end{eqnarray}
\begin{eqnarray}
\nonumber
=I^{a(4)}_{{\rm (CL)}}=\int_{t_{2}}^{t_{1}} dt
A^{a}_{{\rm (CL)}t} |_{x=x_{1}, y=0, z=0}
\end{eqnarray}
\begin{eqnarray}
=\lambda^a (\frac{1}{2}k^{-1/2}_{N_{\rm D}})
[H_{{\rm (CL)}}(t_{2},x_{2})-H_{{\rm (CL)}}(t_{1},x_{2})],
\end{eqnarray}
where $x_{2}=-x_{1}$. From Eqs. (\ref{eqn:hDef})-(\ref{eqn:hant}), it follows
\begin{eqnarray}
\nonumber
H_{{\rm (CL)}}(t,x)=\int dt^{\prime} 
h(t^{\prime},x)
\end{eqnarray}
\begin{eqnarray}
=-\arccos[\exp(-a_{\rm c}xt)],
\end{eqnarray}
where $H(t,x)$ has the property $H(t,x)=-H(t,-x)$, and the summation of the above integrals amounts to
\begin{eqnarray}
\label{eqn:WCL1}
\nonumber
I^{a}_{{\rm (CL)}}
=I^{a(1)}_{{\rm (CL)}}
+I^{a(2)}_{{\rm (CL)}}
+I^{a(3)}_{{\rm (CL)}}
+I^{a(4)}_{{\rm (CL)}}
\end{eqnarray}
\begin{eqnarray}
=-\lambda^a (k^{-1/2}_{N_{\rm D}})\{ \arccos[\exp(-a_{\rm c}x_{2}t_{2})]-\arccos[\exp(-a_{\rm c}x_{
2
}t_{1})] \}.
\end{eqnarray}
The last term $\arccos[\exp(-a_{\rm c}x_{
2
}t_{1})]$ can be neglected for small $a_{\rm c}x_{
2
}t_{1}$.
\par
In contrast, regarding the Wilson loop of the classical unlocalized function, the $x$-component of the vector potential becomes $A^{a}_{{\rm (CU)}x}(t,{\bf x})=0$ from Eqs. (\ref{eqn:Rhox}), (\ref{eqn:CUF1}) and (\ref{eqn:CSo21}), resulting in the vanishing of the following line integrals along the $x$-axis as
\begin{eqnarray}
\label{eqn:WUL1}
\nonumber
I^{a(1)}_{{\rm (CU)}}=
\int_{x_{1}}^{x_{2}} dx A^{a}_{{\rm (C
U
)}x} |_{t=t_{1}, y=0, z=0}
\end{eqnarray}
\begin{eqnarray}
=I^{a(3)}_{{\rm (CU)}}=
\int_{x_{2}}^{x_{1}} dx A^{a}_{{\rm (
C
U
)}x} |_{t=t_{2}, y=0, z=0} = 0.
\end{eqnarray}
Additionally, using Eq. (\ref{eqn:CUF2}), which states that the $t$-component of the classical unlocalized function is equal to zero at $y=0$, we have
\begin{eqnarray}
\label{eqn:WUL2}
\nonumber
I^{a(2)}_{{\rm (CU)}}
=\int_{t_{1}}^{t_{2}} dx A^{a}_{{\rm (C
U
)}t} |_{x=x_{1}, y=0, z=0}
\end{eqnarray}
\begin{eqnarray}
=I^{a(4)}_{{\rm (CU)}}
=\int_{x_{1}}^{x_{2}} dx A^{a}_{{\rm (C
U
)}t} |_{x=x_{2}, y=0, z=0} = 0.
\end{eqnarray}
From Eqs. (\ref{eqn:WUL1}) and (\ref{eqn:WUL2}), the total contribution of the classical unlocalized function thus vanishes as
\begin{eqnarray}
\label{eqn:ULAL}
I^{a}_{{\rm (CU)}}=
 I^{a(1)}_{{\rm (CU)}}
+I^{a(2)}_{{\rm (CU)}}
+I^{a(3)}_{{\rm (CU)}}
+I^{a(4)}_{{\rm (CU)}}
=0,
\end{eqnarray}
and considering Eq. (\ref{eqn:WCL1}) we obtain
\begin{eqnarray}
\label{eqn:CLAL}
I^{a}_{{\rm (C)}}=I^{a}_{{\rm (CL)}}+I^{a}_{{\rm (CU)}}
=I^{a}_{{\rm (CL)}}.
\end{eqnarray}
\par
The interaction potential between the fermion particle and antiparticle is derived from the Wilson loop using Lie algebra.
Considering the properties of the Lie matrices $T^{a}$
\begin{eqnarray}
{\rm Tr}( T^{a}T^{b} )=\frac{1}{2}\delta_{a,b},
\end{eqnarray}
\begin{eqnarray}
{\rm Tr}(T^{a})=0,
\end{eqnarray}
the normalization constant $k_{N_{\rm D}}$ is set to satisfy
\begin{eqnarray}
k^{-1}_{N_{\rm D}}g_{\rm c}^2 
\sum_{a=1}^{3}\sum_{b=1}^{3} \lambda^a\lambda^b T^{a}T^{b}
=\sum_{a=1}^{3} (k^{-1/2}_{N_{\rm D}}g_{\rm c}\lambda^a)^2 T^{a}T^{a}
=I^{(3)},
\end{eqnarray}
where $I^{(3)}$ is the unit matrix for $2 \times 2$ submatrices, such as SU(2), embedded in the $N \times N$ matrix with each element being equal to zero.
In the above relation, let
\begin{eqnarray}
\xi^a=k^{-1/2}_{N_{\rm D}}g_{\rm c}\lambda^a,
\end{eqnarray}
then, from
\begin{eqnarray}
\cos(y)=
\sum_{n=0}^{\infty} (-1)^{n} \frac{y^{2n}}{(2n)!},
\end{eqnarray}
\begin{eqnarray}
\sin(y)=
\sum_{n=0}^{\infty} (-1)^{n} \frac{y^{2n+1}}{(2n+1)!},
\end{eqnarray}
we obtain
\begin{eqnarray}
\nonumber
  {\rm Tr}[\exp(iy^{\prime}\sum_{a=1}^{3} \xi^a T^a)]
= {\rm Tr}[\cos( y^{\prime}\sum_{a=1}^{3} \xi^a T^a)]
+i{\rm Tr}[\sin( y^{\prime}\sum_{a=1}^{3} \xi^a T^a)]
\end{eqnarray}
\begin{eqnarray}
={\rm Tr}[\cos(y^{\prime}I^{(3)})]
=[{\rm Tr}(I^{(3)})]\cos(y^{\prime}).
\end{eqnarray}
Considering Eq. (\ref {eqn:WCL1}), we put $y^{\prime }
=\arccos[\exp(-a_{\rm c}x_{2}t_{2})]
$ in the above relation. Then, using Eqs. (\ref{eqn:WILX})-(\ref{eqn:CLAL}), the classical Wilson loop in Eq. (\ref{eqn:CWLP}) becomes
\begin{eqnarray}
\nonumber
W_{\rm C}=[{\rm Tr}(I^{(3)})]
\biggl \{ \cos \{ \arccos[\exp(-a_{\rm c}x_{2}t_{2})] \} \biggr \}
\end{eqnarray}
\begin{eqnarray}
\nonumber
=[{\rm Tr}(I^{(3)})]
\biggl \{ \cos \{ \cos^{-1}[\exp(-a_{\rm c}x_{2}t_{2})] \} \biggr \}
\end{eqnarray}
\begin{eqnarray}
\nonumber
=\exp [-a_{\rm c}x_{2}(t_{2} -t_{1})+\ln[ {\rm Tr}(I^{(3)})] ]
\end{eqnarray}
\begin{eqnarray}
=\exp[-\frac{a_{\rm c}}{2}(x_{2}-x_{1})(t_{2} -t_{1})+\ln[ {\rm Tr}(I^{(3)})]],
\end{eqnarray}
where we have used $x_{2}=-x_{1}$.
Note that this area law is not obtained for the Abelian gauge fields, because the matrices $T^a$ are not involved in the Abelian gauge fields. From this relationship, we derive the linear potential
\begin{eqnarray}
\label{eqn:CWL}
\nonumber
V_{\rm C}=-\frac{\ln(W_{\rm C})}{t_{2} -t_{1}}
\end{eqnarray}
\begin{eqnarray}
=\frac{a_{\rm c}}{2}(x_{2}-x_{1}).
\end{eqnarray}
\par
Here, we provide physical implications of the confining classical localized function. The localized function $A_{(\rm C)\mu}$, which is composed of the soliton-like function $h(t, x)$, is related to the soliton solutions of the sine-Gordon equation \cite{PerSky} given by
\begin{eqnarray}
\phi_{-}(t,x)=4 \arctan [ \exp(-x) ],
\end{eqnarray}
\begin{eqnarray}
\phi_{+}(t,x)=4 \arctan [ \exp(+x) ].
\end{eqnarray}
The function $H_{{\rm (CL)}}(t,x)$ is derived by integrating $h(t,x)$, which composes the classical confining localized function $A_{{\rm ({\rm CL})} \mu}$. (The function $A_{{\rm ({\rm CL})} \mu}$ is part of the classical solution $A_{(\rm C)\mu}$.) This $H_{{\rm (CL)}}(t,x)$ for $t=1$ and $a_{\rm c}=1$ is related to the solution of the sine-Gordon equation as
\begin{eqnarray}
H_{{\rm (CL)}}(1,x)
=-\arccos[ \tan \frac{1}{4}\phi_{-}(t,x)  ]
\hspace{2ex}
\mbox{ for $x \geq 0$},
\end{eqnarray}
\begin{eqnarray}
H_{{\rm (CL)}}(1,x)
=
\hspace{2ex}
\arccos[ \tan \frac{1}{4}\phi_{+}(t,x) ]
\hspace{2ex}
\mbox{ for $x < 0$}.
\end{eqnarray}
\par

\subsection{Scale-invariant energy of the non-Abelian gauge field and string tension of the linear potential and Polyakov's confining energy \label{sec:Sec43}}

In the case of an atom, the characteristic scale is the Bohr radius, which is of the order of $10^{-10}$ m. The Bohr radius is a function of
Planck's
constant, the electron mass and coupling constant of quantum electrodynamics (QED). The electron mass and coupling constant therefore determine the Bohr radius. In contrast, non-Abelian gauge fields have a somewhat different intrinsic scale-invariant energy (length/time) and describe quantum chromodynamics (QCD). This scale-invariant energy is a given fundamental constant similar to
Planck's
constant and cannot be determined by theory. Thus, the solutions of the non-Abelian Yang-Mills equation must be consistent with this scale-invariance. This subsection examines such specific scale-invariant phenomena in the scheme of the non-Abelian gauge field.
\par
We now evaluate the dominant energy of the confining classical field, which is a soliton-like localized function (object; wave packet). Using the energy tensor, the energy of the solitonlike object at Minkowski time $t^{\prime}$ (corresponding to Euclidean time $t$) is calculated by
\begin{eqnarray}
E_{{\rm (CL)}}=\int dxdydz T_{00{\rm (CL)}},
\end{eqnarray}
where
\begin{eqnarray}
T_{00{\rm (CL)}}=  \frac{\partial L_{{\rm F}{\rm (CL)}}}
{\partial(\frac{\partial A_{{\rm (CL)}\nu}}{\partial x^{\prime}_{0}})}
\frac{\partial A_{{\rm (CL)}\nu}}{\partial x^{\prime}_{0}} -L_{{\rm F}{\rm (CL)}},
\end{eqnarray}
and
\begin{eqnarray}
L_{{\rm F}{\rm (CL)}}=-\frac{2}{4} {\rm Tr} (F_{{\rm (CL)}\mu\nu})^2.
\end{eqnarray}
By integrating the function, the following energy of the solitonlike localized function at time $t^{\prime}$ is obtained
\begin{eqnarray}
\nonumber
E_{{\rm (CL)}}(\epsilon_{t^{\prime}})
=\{ 2 \int_{0}^{\infty} dx [\frac{1}{g_{\rm c}
}h(\epsilon_{t^{\prime}},x)]^2 \}
\end{eqnarray}
\begin{eqnarray}
\times
\frac{4}{2}\{ \int_{0}^{\infty} \int_{0}^{\infty} dydz 
[(\partial_{y} w(y)w(z))^2+
(w(y)\partial_{z}w(z))^2] \}.
\end{eqnarray}
This integral is independent of $d$ (the thickness of the localized function), and we have taken the limits as $\epsilon_{y} \rightarrow 0$, $\epsilon_{z} \rightarrow 0$ and $\epsilon_{x} \rightarrow 0$. Using Eq. (\ref{eqn:hDef}), the above integral yields
\begin{eqnarray}
E_{{\rm (CL)}}(\epsilon_{t^{\prime}})
=\frac{\zeta(3)\Gamma(3)}{4 g_{\rm c}^2}
\frac{1}{a_{\rm c}(\epsilon_{t^{\prime}})^3},
\end{eqnarray}
where $\zeta$ and $\Gamma$ are the zeta and  gamma functions, respectively.
\par
We then considered the scale-invariant energy relationship for the linear potential. This relationship at the scale-invariant time $\epsilon^{\prime}$ is such that the above energy of the solitonlike object is equal to $E_{(\rm LP)}(\epsilon_{t^{\prime}})$, which is the energy decrease of the particle-antiparticle pair due to the linear potential, and from Eq. (\ref{eqn:CWL}) we have
\begin{eqnarray}
E_{{\rm (CL)}}(\epsilon_{t^{\prime}})=E_{(\rm LP)}(\epsilon_{t^{\prime}})=|-a_{\rm c}\epsilon_{t^{\prime}}|,
\end{eqnarray}
with speed of light set to unity, $c=1$. The symbol $2\epsilon_{t^{\prime}}$ represents the maximum effective particle-antiparticle distance at the scale-invariant time $t^{\prime}$.
(Equations (\ref{eqn:CFEQ1}) and (\ref{eqn:CFEQ2}) state that after the scale-invariant time, some energy of the solitonlike object is transferred to the field of the unlocalized function via the time-dependent source charge of the field.)
The parameter of the classical field $a_{\rm c}$ is determined from the above equation. 
By letting the maximum effective radius of the fundamental particle $R_{\rm p}=\epsilon_{t^{\prime}}$ at the scale-invariant time $t^{\prime}$, the maximum effective particle-antiparticle distance (effective diameter $D_{\rm p}$ of the fundamental particle) is the light cone diameter at time $\epsilon_{t^{\prime}}$, that is,
\begin{eqnarray}
D_{\rm p}=2R_{\rm p}=2\epsilon_{t^{\prime}}.
\end{eqnarray}
Meanwhile, the continuum scale-invariant length, $\lambda_{\rm MOM}$, which is the inverse of the energy, is the following size of the particle-antiparticle pair (diameter of the pair)
\begin{eqnarray}
\lambda_{\rm MOM}=2D_{\rm p}.
\end{eqnarray}
Consequently, from Eq. (\ref{eqn:CWL}), the string tension $\sigma$ amounts to
\begin{eqnarray}
\label{eqn:StSIE}
\sigma=\frac{1}{2} a_{\rm c}
=\frac{4\zeta(3)^{1/2}\Gamma(3)^{1/2}\lambda_{\rm MOM}^2}
{g_{\rm c}}.
\end{eqnarray}
\par
As mentioned in Subsection \ref{sec:Sec63} in detail, the above theoretical relation (with the use of the classical coupling constant \cite{MPes}) reproduces the experimental string tension observed from the Regge trajectory \cite{ColReg, Naga} in the form $1/(4\sigma)=0.93$ GeV$^{-2}$, that is, $\sqrt{\sigma}=518.5$ MeV at $\Lambda_{\rm MOM}=229.9$ MeV, which is consistent with the observed QCD scale-invariant energy of around 213 MeV \cite{MPes}.
\par
Now, the binding energy between a fermion particle and antiparticle at finite temperatures is evaluated using the Polyakov line \cite{Polya}, which corresponds to the Wilson loop. We introduce $\tau=1/k_{B}T$, where $k_{B}$ and $T$ are the Boltzmann constant and temperature, respectively. By replacing $t$ by $\tau^{\prime}$ and imposing the periodic condition along the temperature axis, we obtain the classical localized function at finite temperatures
\begin{eqnarray}
\nonumber
A^{a}_{{\rm (CL)}t}(t,x,y,z) \rightarrow A^{a}_{{\rm (CL)}\tau}(\tau^{\prime},x,y,z) 
\end{eqnarray}
\begin{eqnarray}
=A^{a}_{{\rm (CL)}t}(\tau^{\prime},x,y,z)+A^{a}_{{\rm (CL)}t}(\tau-\tau^{\prime},x,y,z).
\end{eqnarray}
The Polyakov line is the line integral
\begin{eqnarray}
P_{\tau}=
{\rm Tr}\{
\exp[-ig_{\rm c} \sum_{a}(
\int_{\tau_{\epsilon}}^{\tau-\tau_{\epsilon}} d\tau^{\prime}
A^{a}_{{\rm (CL)}\tau} |_{x=x_{2}, y=0, z=0} 
T^{a})]
\},
\end{eqnarray}
where the start point can be regarded as the end point ($\tau_{\epsilon}$ is an infinitesimally small quantity. Using the unitary gauge transformation for a general physical quantity $U^{-1}A_{\rm ph}U=A_{\rm ph}$, the Polyakov line is gauge invariant and leads to
\begin{eqnarray}
\label{eqn:PolyL}
\nonumber
P_{\tau}=\cos\{\arccos[\exp(-\sigma r\tau)]
 -\arccos[\exp(-\sigma r\tau_{\epsilon})]\}
\end{eqnarray}
\begin{eqnarray}
\nonumber
\approx \cos\{\arccos[\exp(-\sigma r\tau)]\}
=\cos\{\cos^{-1}[\exp(-\sigma r\tau)]\}
\end{eqnarray}
\begin{eqnarray}
=\exp(-\sigma r\tau),
\end{eqnarray}
where $\tau =1/(k_{B}T)$ ($k_{\rm B}$ and $T$ are the Boltzmann constant and temperature, respectively).
Then,
\begin{eqnarray}
\label{eqn:EPOLA}
\epsilon_{\rm q}=-\ln(P_{\tau}) \approx -\ln[\exp(\frac{-E_{\rm B}}{k_{\rm B}T})]=\frac{E_{\rm B}}{k_{\rm B}T}
=\frac{\sigma r}{k_{\rm B}T},
\end{eqnarray}
with $\epsilon_{\rm q}$ being the binding energy of the fundamental fermion and antifermion  pair, and
$E_{\rm B}=\sigma r$.
Equation (\ref{eqn:EPOLA}) indicates that confinement occurs for large $\tau$ at low temperatures and deconfinement occurs for small $\tau$ at high temperatures. This binding energy is quite different from that of the superconducting pair, which vanishes above the superconducting critical temperature. It is experimentally reported that a quark and gluon plasma (QGP) at high temperatures may be a fluid with vanishing viscosity.
\par

\section{Quantum field in path integral around classical field as a vacuum \label{sec:Sec5}}

\subsection{Gauge invariance in the non-Abelian gauge field scheme with fields expressed in terms of step-function-type basis functions \label{sec:Sec51}}

Now, we present a formalism of the quantum fluctuations for non-Abelian gauge fields around the classical field as a vacuum in four-dimensional spacetime. The quantum field is expressed in terms of step-function-type basis functions and their derivatives with finite degrees of freedom to construct the gauge-invariant scheme in the spacetime continuum. We set the parameter spacetime in Subsection \ref{sec:Sec22} to be identical to real spacetime for convenience, and we use the basis functions around the central point, $x_{p}=(t_{(k,l,m,n)}, {\bf x}_{(k,l,m,n)})$ of the hypercube and its derivative, given by Eqs.
(\ref{eqn:DeEB})-(\ref{eqn:DB3P}).
By replacing the parameter coordinates $x_{\mu{\rm P}}$ by real coordinates  $x_{\mu}$, the basis functions become
\begin{eqnarray}
\label{eqn:BSE4}
\Omega^{4\tilde{\rm E}}_{p}(x)=\Omega^{4 \tilde{\rm E}}_{(k,l,m,n)}(t,x,y,z)=\Omega^{\tilde{\rm E}}_{k}(t)\Omega^{\tilde{\rm E}}_{l}(x)\Omega^{\tilde{\rm E}}_{m}(y)\Omega^{\tilde{\rm E}}_{n}(z),
\end{eqnarray}
\begin{eqnarray}
\label{eqn:BSE3}
\Omega^{\delta 3\tilde{\rm E} }_{\mu p}(x)
=\Omega^{\delta 3\tilde{\rm E}-}_{\mu p}(x)-\Omega^{\delta 3\tilde{\rm E}+}_{\mu p}(x),
\end{eqnarray}
where the above  symbols such as $\Omega^{\delta 3\tilde{\rm E}-}_{t p}(x)$ are represented (in an example case for a function of $t$) by
\begin{eqnarray}
\nonumber
\Omega^{\delta 3\tilde{\rm E}-}_{t p}(x)=\Omega^{\delta 3\tilde{\rm E}-}_{t (k,l,m,n)}(t,x,y,z)
\end{eqnarray}
\begin{eqnarray}
=\partial_{t}\Omega^{\tilde{\rm E}}_{k}(t)|_{t=t_{k-1/2}}
\Omega^{\tilde{\rm E}}_{l}(x)\Omega^{\tilde{\rm E}}_{m}(y)\Omega^{\tilde{\rm E}}_{n}(z)\Delta.
\end{eqnarray}
\begin{eqnarray}
\nonumber
\Omega^{\delta 3\tilde{\rm E}+}_{t p}(x)=\Omega^{\delta 3\tilde{\rm E}+}_{t (k,l,m,n)}(t,x,y,z)
\end{eqnarray}
\begin{eqnarray}
=-\partial_{t}\Omega^{\tilde{\rm E}}_{k}(t)|_{t=t_{k
+
1/2}}
\Omega^{\tilde{\rm E}}_{l}(x)\Omega^{\tilde{\rm E}}_{m}(y)\Omega^{\tilde{\rm E}}_{n}(z)\Delta.
\end{eqnarray}
The basis function $\Omega^{4\tilde{\rm E}}_{p}(x)$ has the property that
\begin{eqnarray}
\Omega^{4\tilde{\rm E}}_{p}(x)\Omega^{4\tilde{\rm E}}_{q}(x)=\delta_{p,q}\Omega^{4\tilde{\rm E}}_{p}(x).
\end{eqnarray}
\par
Here, the Dirac field with its phase and the gauge field are denoted, in terms of basis functions $\Omega^{\tilde{\rm E}4}_{q}(x)$,
$\Omega^{\delta 3\tilde{\rm E}
-
}_{\mu p}(x)$ 
and 
$\Omega^{\delta 3\tilde{\rm E}
+
}_
{\mu p}
(x)$,
as
\begin{eqnarray}
\label{eqn:GEx1}
\psi(x)=\sum_{p}
\psi_{p}\Omega^{4\tilde{\rm E}}_{p}(x)
\exp [i \alpha^{0}(x)+ i\sum_{a}\alpha^{a}(x) T^{a}],
\end{eqnarray}
\begin{eqnarray}
\label{eqn:GEx2}
\alpha^{a}(x)=\sum_{p}\alpha^{a}_{p}\Omega^{4\tilde{\rm E}}_{p}(x)
=\sum_{p}\alpha^{\prime a}_{p}\Omega^{4\tilde{\rm E}}_{p}(x)\Delta,
\end{eqnarray}
\begin{eqnarray}
\label{eqn:GEx3}
A^{a}_{\mu}(x)=\sum_{p}[A^{a}_{({\rm P}1)\mu p}\Omega^{4\tilde{\rm E}}_{p}(x)
+A^{a}_{({\rm P}2) p}\Omega^{\delta 3\tilde{\rm E}
-
}_{\mu p}(x)
-A^{a}_{({\rm P}3) p}\Omega^{\delta 3\tilde{\rm E}+}_{\mu p}(x)
].
\end{eqnarray}
The conventional infinitesimal gauge transformation for the local non-Abelian gauge field is given by
\begin{eqnarray}
\label{eqn:GEx4}
\delta A^{a}_{\mu}(x)=-(1/g)\partial_{\mu} \delta\alpha^{a}(x)
-\sum_{b,c} f^{bca}\delta\alpha^{b}(x)A^{c}_{\mu}(x).
\end{eqnarray}
By substituting Eqs. (\ref{eqn:GEx2})-(\ref{eqn:GEx3}) into Eq. (\ref{eqn:GEx4}), we obtain
(it is also expressed in terms of the independent basis functions $\Omega^{\delta 3\tilde{\rm E}-}_
{\mu p}
$ and $\Omega^{\delta 3\tilde{\rm E}+}_
{\mu p}
$ multiplied by the corresponding independent expansion coefficient)
\begin{eqnarray}
\label{eqn:GEx5}
\nonumber
\delta A^{a}_{\mu}(x)
=-\sum_{p} (1/g)\delta\alpha^{\prime a}_{p}
\Omega^{\delta 3\tilde{\rm E}}_{\mu p}(x)
\end{eqnarray}
\begin{eqnarray}
-\sum_{b,c}f^{bca}\sum_{p}
\delta\alpha^{b}_{p}A^{c}_{({\rm P}1) \mu p}\Omega^{4\tilde{\rm E}}_{p} (x)
-\sum_{b,c}f^{bca}\sum_{p}
\delta\alpha^{b}_{p}A^{c}_{({\rm P}2) \mu p}\Omega^{\delta 3\tilde{\rm E}
-
}_
{\mu p}
(x).
\end{eqnarray}
From Eq. (\ref{eqn:GEx3}), we have
\begin{eqnarray}
\label{eqn:GEx6}
\delta A^{a}_{\mu}(x)=
\sum_{p}[\delta A^{a}_{({\rm P}1)\mu p}\Omega^{4\tilde{\rm E}}_{p}(x)
+\delta A^{a}_{({\rm P}2) p}\Omega^{\delta 3\tilde{\rm E}
-
}_{\mu p}(x)
-\delta A^{a}_{({\rm P}3) p}\Omega^{\delta 3\tilde{\rm E}+}_{\mu p}(x)
].
\end{eqnarray}
By equating the coefficients of the basis functions $\Omega^{4\tilde{\rm E}}_{p}(x)$,
$\Omega^{\delta 3\tilde{\rm E}
-
}_
{\mu p}
(x)$ 
and 
$\Omega^{\delta 3\tilde{\rm E}
+
}_
{\mu p}
(x)$
in Eqs. (\ref{eqn:GEx5}) and (\ref{eqn:GEx6}), the following equalities are derived
\begin{eqnarray}
\label{eqn:GIV1}
\delta A^{a}_{({\rm P}1)\mu p}
=
-\sum_{b,c}f^{bca}\delta\alpha^{b}_{p}
A^{c}
_{({\rm P}1) \mu p},
\end{eqnarray}
\begin{eqnarray}
\label{eqn:GIV2}
\nonumber
\delta A^{a}_{({\rm P}2) p}=-(1/g)\delta\alpha^{\prime a}_{p}
-\sum_{b,c}f^{bca}\delta\alpha^{b}_{p}A^{c}_{({\rm P}2) \mu p},
\end{eqnarray}
\begin{eqnarray}
\label{eqn:GIV2}
\delta A^{a}_{({\rm P}3) p}=-(1/g)\delta\alpha^{\prime a}_{p}.
\end{eqnarray}
Consequently, the present formalism is gauge invariant because of the relations of Eqs. (\ref{eqn:GIV1}) and (\ref{eqn:GIV2}). 
The merit of using step functions is the easy evaluation of quantities such as the cross term $\Omega^{4\tilde{\rm E}}_{p}(x)\Omega^{4\tilde{\rm E}}_{q}(x)$, which is equal to $\delta_{p,q}\Omega^{4\tilde{\rm E}}_{p}(x)$, and its integral.
\par
\subsection{Quantum action for non-Abelian gauge fields expanded in terms of step-function-type basis functions \label{sec:Sec52}}

As described in the previous section, the field $A^{a}_{\mu}(x)$ in the present formalism comprises the classical field $A_{{\rm (C)}\mu}(x)$ as a vacuum and the quantum field (fluctuation) $A^{a}_{{\rm (Q)}\mu}(x)$ around the classical field, as denoted by
\begin{eqnarray}
A^{a}_{\mu}(x)=A^{a}_{{\rm (C)}\mu}(x)+A^{a}_{{\rm (Q)}\mu}(x).
\end{eqnarray}
Then, the zeroth order of the action for the non-Abelian gauge fields is the classical field, and the first-order action vanishes due to the classical equations of motion, while the second-order remains. 
We then consider the following quantum action in the Feynman gauge 
\begin{eqnarray}
S=S^{(2)}+S^{(3)}+S^{(4)},
\end{eqnarray}
where
\begin{eqnarray}
\label{eqn:QAc2}
S^{(2)}=\frac{1}{2}\sum_{a} \int d^4x 
(\partial_{\nu} A^{a}_{{\rm (Q)} \mu}\partial_{\nu}A^{a}_{{\rm (Q)} \mu}),
\end{eqnarray}
\begin{eqnarray}
\nonumber
S^{(3)}=-g \sum_{a,b,c} f^{abc} \int d^4x
(A^{b}_{{\rm (Q)} \mu}A^{c}_{{\rm (Q)} \nu}\partial_{\mu}A^{a}_{\rm (C) \nu}
\end{eqnarray}
\begin{eqnarray}
+A^{b}_{{\rm (Q)} \mu}A^{c}_{\rm (C) \nu}\partial_{\mu}A^{a}_{{\rm (Q)} \nu}
+A^{b}_{\rm (C) \mu}A^{c}_{{\rm (Q)} \nu}\partial_{\mu}A^{a}_{{\rm (Q)} \nu}
+A^{b}_{\rm (Q) \mu}A^{c}_{{\rm (Q)} \nu}\partial_{\mu}A^{a}_{{\rm (Q)} \nu}),
\end{eqnarray}
\begin{eqnarray}
\nonumber
S^{(4)}=
\nonumber
\frac{g^{2}}{4} \sum_{a,b,c,d,e} f^{abc}f^{ade} \int d^4x
(A^{b}_{{\rm (Q)} \mu} A^{c}_{{\rm (Q)} \nu} A^{d}_{\rm (C) \mu} A^{e}_{\rm (C) \nu}
\end{eqnarray}
\begin{eqnarray}
\nonumber
+A^{b}_{{\rm (Q)} \mu} A^{c}_{\rm (C) \nu} A^{d}_{{\rm (Q)} \mu} A^{e}_{\rm (C) \nu}
+A^{b}_{{\rm (Q)} \mu} A^{c}_{\rm (C) \nu} A^{d}_{\rm (C) \mu} A^{e}_{{\rm (Q)} \nu}
\end{eqnarray}
\begin{eqnarray}
\nonumber
+A^{b}_{\rm (C) \mu} A^{c}_{{\rm (Q)} \nu} A^{d}_{{\rm (Q)} \mu} A^{e}_{\rm (C) \nu}
+A^{b}_{\rm (C) \mu} A^{c}_{{\rm (Q)} \nu} A^{d}_{\rm (C) \mu} A^{e}_{{\rm (Q)} \nu}
\end{eqnarray}
\begin{eqnarray}
\nonumber
+A^{b}_{\rm (C) \mu} A^{c}_{\rm (C) \nu} A^{d}_{{\rm (Q)} \mu} A^{e}_{{\rm (Q)} \nu}
+A^{b}_{\rm (Q) \mu} A^{c}_{\rm (Q) \nu} A^{d}_{{\rm (Q)} \mu} A^{e}_{{\rm (C)} \nu}
\end{eqnarray}
\begin{eqnarray}
\nonumber
+A^{b}_{\rm (Q) \mu} A^{c}_{\rm (Q) \nu} A^{d}_{{\rm (C)} \mu} A^{e}_{{\rm (Q)} \nu}
+A^{b}_{\rm (Q) \mu} A^{c}_{\rm (C) \nu} A^{d}_{{\rm (Q)} \mu} A^{e}_{{\rm (Q)} \nu}
\end{eqnarray}
\begin{eqnarray}
+A^{b}_{\rm (C) \mu} A^{c}_{\rm (Q) \nu} A^{d}_{{\rm (Q)} \mu} A^{e}_{{\rm (Q)} \nu}
+A^{b}_{\rm (Q) \mu} A^{c}_{\rm (Q) \nu} A^{d}_{{\rm (Q)} \mu} A^{e}_{{\rm (Q)} \nu}).
\end{eqnarray}
The above quantum action comprises kinetic and self-interaction terms and the cubic and quartic terms contain the classical field.
The cubic term is small because the quantum coupling constant $g$ is small for short small lattice spacings, and the quartic term is smaller than the cubic term for such small lattice spacings.
\par
The quantum field is expressed in terms of basis functions $\Omega^{4\tilde{\rm E}}_{p}(x)$,
$\Omega^{\delta 3\tilde{\rm E}-}_{\mu p}(x)$ 
and 
$\Omega^{\delta 3\tilde{\rm E}+}_{\mu p}(x)$
in Eqs. (\ref{eqn:BSE4}) and (\ref{eqn:BSE3}), respectively, and the contribution of quantum fluctuations is evaluated using the path integral. Owing to the gauge invariance of the present formalism, the coefficients of the basis set of
$\Omega^{\delta 3\tilde{\rm E}-}_{\mu p}(x)$ 
and 
$\Omega^{\delta 3\tilde{\rm E}+}_{\mu p}(x)$
can be canceled by the gauge transformation to reduce the field to the following form with only the quantum fluctuations:
\begin{eqnarray}
A^{a}_{{\rm (Q)} \mu}(x)=\sum_{p} A^{a}_{({\rm Q}) \mu p} \Omega^{4\tilde{\rm E}}_{p}(x).
\end{eqnarray}
\par
Here, we consider the local mass term in the action  for the limit as $d \rightarrow 0$ (the quantity $d$ given in Eq. (\ref{eqn:wyzD}) is the sheet thickness of the classical soliton-like object described by the classical localized function)
\begin{eqnarray}
S^{(41)}_{\rm (CQ)}=
\frac{1}{4}g^2f^{abc}f^{ade} \int d^4x
A^{b}_{{\rm (Q)} \mu} A^{c}_{{\rm (C)}t} A^{e}_{{\rm (C)} t} A^{d}_{{\rm (Q)} \mu}.
\end{eqnarray}
When $b=d$ and $
c
=e$ in the above equation, we have
\begin{eqnarray}
\nonumber
S^{(41)}_{\rm (CQ)}=
\frac{1}{4}g^2f^{abc}f^{abc} \int d^4x
A^{b}_{{\rm (Q)} \mu} A^{c}_{{\rm (C)}t} A^{c}_{{\rm (C)}t} A^{b}_{{\rm (Q)} \mu}
\end{eqnarray}
\begin{eqnarray}
\label{eqn:mass}
=\frac{1}{4}g^2(f^{abc})^2 \int d^4x
A^{b}_{{\rm (Q)} \mu} (A^{c}_{{\rm (C)}t})^2 A^{b}_{{\rm (Q)} \mu}.
\end{eqnarray}
The above equation states that the following quantity
\begin{eqnarray}
\label{eqn:MasGp}
m_{\rm loc}=\frac{1}{4} g^2 (f^{abc})^2 (A^{c}_{{\rm (C)}t})^2
\hspace{2ex}\mbox{ with } m_{\rm loc} > 0,
\end{eqnarray}
which is multiplied by two identical quantum fields
$A^{b}_{{\rm (Q)} \mu}$ and $A^{b}_{{\rm (Q)} \mu}$ in Eq. (\ref{eqn:mass}),
implies a local mass (gap)
$((2m_{\rm loc})^{1/2})$
with the non-zero positive value for non-zero structure constants $f^{abc}$ in Eqs. (\ref{eqn:SCN2}) and (\ref{eqn:SCN3}).
\par
Now, the action of the kinetic terms in Eq. (\ref{eqn:QAc2}), denoted as $S^{(2)a}_{\rm Q,Q}$, is expressed in terms of the basis functions to give
\begin{eqnarray}
\nonumber
S^{(2)a}_{\rm Q,Q}=\frac{1}{2}\int d^4x 
(\partial_{\nu}\sum_{p}A^{a}_{({\rm Q}) \mu p}\Omega^{4{\tilde{\rm E}}}_{p})
(\partial_{\nu}\sum_{q}A^{a}_{({\rm Q}) \mu q}\Omega^{4{\tilde{\rm E}}}_{q}),
\end{eqnarray}
\begin{eqnarray}
=\sum_{p,q}A^{a}_{({\rm Q})\mu p}M^{{\rm Q},{\rm Q}}_{pq} A^{a}_{({\rm Q})\mu q},
\end{eqnarray}
where
\begin{eqnarray}
M^{{\rm Q},{\rm Q}}_{pq}
=\frac{1}{2}\int d^4x 
(\partial_{\nu}\Omega^{4 \tilde{\rm E}}_{p})
(\partial_{\nu}\Omega^{4 \tilde{\rm E}}_{q}).
\end{eqnarray}
In a similar way to Eqs. (\ref{eqn:SEq01})-(\ref{eqn:SEq20}), the above action of the kinetic terms is diagonalized to yield the eigenvalues
\begin{eqnarray}
\eta^{av}_{({\rm Q})(0)}=[(1-c_{0})+(1-c_{1})+(1-c_{2})+(1-c_{3})]
\Delta^{2},
\end{eqnarray}
where
\begin{eqnarray}
c_{\mu}=\cos (z_{\mu}),
\hspace{2ex}
z_{\mu}=\frac{j_{\mu}\pi}{N_{(\mu)}+1}.
\end{eqnarray}
\par

\subsection{Quantum Wilson loop \label{sec:Sec53}}

We next consider the contribution of the quantum fluctuations to the Wilson loop, neglecting higher-order contributions due to asymptotic freedom (small coupling). The quantum action we consider is
\begin{eqnarray}
\label{eqn:QQTM2}
S_{\rm Q}=\sum_{a}\sum_{p,q} M^{{\rm Q},{\rm Q}}_{pq}
A^{a}_{{\rm (Q)}\mu p}A^{a}_{{\rm (Q)}\mu q},
\end{eqnarray}
and by using the classical Wilson loop $W_{\rm C}$, the $2 \times 2$ unit submatrix $I^{(3)}$ embedded in the $N \times N$ matrix with vanishing components, and the basis set $\Omega^{4 \tilde{\rm E}}_{p}$ in Eq. (\ref{eqn:BSE4}), the Wilson loop in the path integral can be given by
\begin{eqnarray}
W_{\rm Q}={\rm Tr}\biggl \{
W_{\rm C}I^{(3)}
\frac{1}{Z_{\rm N}} \int D[A^{a}_{{\rm (Q)}\mu p}]
\exp(-S_{\rm Q}) \exp(C)
\biggr \}.
\end{eqnarray}
Here,
\begin{eqnarray}
Z_{\rm N}= \int D[A^{a}_{{\rm (Q)}\mu p}] \exp(-S_{\rm Q}),
\end{eqnarray}
is a normalization constant in the path integral with respect to $A^{a}_{{\rm (Q)}\mu p}$ and
\begin{eqnarray}
C=-ig \sum_{a}\sum_{p} \beta_{\mu p}
A^{a}_{{\rm (Q)}\mu p}T^{a},
\end{eqnarray}
with
\begin{eqnarray}
\beta_{\mu p} = \oint dx_{\mu} \Omega^{4 \tilde{\rm E}}_{p}.
\end{eqnarray}
In the above expression, the pre-factor is the contribution from the classical field, and the quantum fluctuations $A^{a}_{{\rm (Q)}\mu p}$ associated  with the Lie matrices $T^{a}$ are those for the non-Abelian gauge field such as SU($3$).
\par
By the diagonalization of Eq. (\ref{eqn:QQTM2}) using a diagonal matrix $R_{pq}$ associated with the eigenvalues $\eta_{(\mu)p}$
\begin{eqnarray}
S_{\rm Q}=\sum_{a}\sum_{p} \eta_{(\mu)p}
(A^{\prime a}_{{\rm (Q)}\mu p})^{2},
\end{eqnarray}
with
\begin{eqnarray}
A^{\prime a}_{{\rm (Q)}\mu p}=
R_{pq}
A^{a}_{{\rm (Q)}\mu q},
\end{eqnarray}
we obtain
\begin{eqnarray}
\nonumber
W_{\rm Q}={\rm Tr}
\biggl \{ W_{\rm C}I^{(3)} \frac{1}{Z_{\rm N}} \int D[A^{\prime a}_{{\rm (Q)}\mu p}]
\exp(-i \sum_{a} \sum_{p} B_{\mu p}^{\prime a}
A_{{\rm (Q)}\mu p}^{\prime a})
\end{eqnarray}
\begin{eqnarray}
\times
\exp(- \sum_{a} \sum_{p}
\eta_{(\mu)p}(A^{\prime a}_{{\rm (Q)}\mu p})^{2})
\biggr \},
\end{eqnarray}
where
\begin{eqnarray}
B_{\mu p}^{\prime a}= g \sum_{q}
\beta_{\mu q}R^{-1}_{qp} T^{a}.
\end{eqnarray}
\par
The integrals yield Gaussian integrals, and the odd term vanishes independently of the gauge group. We then have
\begin{eqnarray}
\nonumber
W_{\rm Q}={\rm Tr}
\biggl \{
W_{\rm C}I^{(3)} \frac{1}{Z_{\rm N}}\Pi_{a}\Pi_{p}\int
dA^{\prime a}_{{\rm (Q)}\mu p}
\end{eqnarray}
\begin{eqnarray}
\times
\cos(B^{\prime a}_{\mu p} A^{\prime a}_{{\rm (Q)}\mu p})
\exp[-\eta_{(\mu)p}(A^{\prime a}_{{\rm (Q)}\mu p})^{2}]
\biggr \}.
\end{eqnarray}
By using
\begin{eqnarray}
Z_{\rm N}=\Pi_{a}\Pi_{p} ( \frac{\pi}{ \eta_{(\mu)p} })^{1/2},
\end{eqnarray}
and, by writing as ${B}^{\prime a}_{\mu p}= \tilde{B}^{\prime}_{\mu p}T^{a}$, we derive
\begin{eqnarray}
W_{\rm Q}={\rm Tr}
\biggl \{
W_{\rm C} I^{(3)} \exp \{-\sum_{a}[ \sum_{p}
\frac{(\tilde{B}^{\prime}_{\mu p})^{2}}
{4\eta_{(\mu)p}}] T^{a}T^{a} \}
\biggr \}.
\end{eqnarray}
The sum of the Lie matrices is proportional to the unit matrix, which is independent of the gauge group, and the quantum potential implies a Coulomb potential. The analytical confining potential derived above is composed of a classical linear term plus quantum Coulomb term denoted by
\begin{eqnarray}
V(x_{2}-x_{1})
=\frac{a_{\rm c}}{2}(x_{2}-x_{1})+V_{\rm C}(x_{2}-x_{1}).
\end{eqnarray}
In the previous section, the binding energy of the particle and antiparticle at finite temperatures was derived from the Polyakov line in Eq. (\ref{eqn:PolyL}). The quantum fluctuations at finite temperatures are also independent of the gauge group, and the quantum contribution from the non-Abelian case is essentially identical to that from the Abelian case.
\par
The confinement mechanism mentioned in this article also works for the pure Yang-Mills case, where source-antisource creation occurs and the confining potential between the source-antisource is produced. Furthermore, the non-zero classical field generates the local mass (gap) with non-zero positive values for the non-Abelian gauge field as shown in Eq. (\ref{eqn:MasGp}). We note that the mass generation in the pure Yang-Mills field occurs both from the binding between the source-antisource and from the non-zero classical field, as in Eq. (\ref{eqn:MasGp}). This prevents the gauge field, whose energy is less than the binding energy between the source-antisource, from passing through the meaningful region of the gauge fields. This is an answer to the important field-theoretic question by Pauli of whether non-Abelian gauge theory can describe the real physical field.

\section{Energies of a bound fundamental fermion-antifermion pair \label{sec:Sec6}}

\subsection{Experimental Regge trajectory and relativistic classical approach toward the quantum description\label{sec:Sec61}}

Experimentally, a fundamental fermion and antifermion pair reveals a remarkable phenomenon, called the Regge trajectory \cite{ColReg}. This trajectory indicates that the squared system energy, which is the total mass of an object comprising a fermion and antifermion, is proportional to the angular momentum. Some theoretical approaches were performed from field theory and lattice gauge theory, as can be observed in previous literature \cite{KGW74a,KGW74b,Kog79,Kog83,Creu80,Creu83,Drou78,Drou83,APHa,HRot,Card,Baza,Call82,Call83,Call88,Wegn,Rebbi,CreuQ,Montv,Makee,Smit,DeGra06,DeGra10,Weisz}.
A relativistic classical mechanical Hamiltonian \cite{Miya,Maki}, which is comprised of the rotational kinetic energy and confining linear potential reproduces the principal properties of the Regge trajectory.
The kinetic energy in the classical Hamiltonian is not deduced from the theory of Dirac fields, and the relationship between the classical angular momentum and the quantum angular momentum is not clear. Moreover, the linear potential has no basis in the non-Abelian gauge field.
In contrast, attempts by other authors \cite{Kabu,AbeF,Tez91,Tez13} to analytically solve the Dirac equation with a linear potential may not succeed in systematically reproducing the Regge trajectory \cite{ColReg}.
Other theoretical/numerical approaches have not gained clear answers at the quantum level to the mechanism and origin of the large binding energy (mass) of the paired fermions compared with the masses of the composite fermion and antifermion.
\par
Conversely, the present approach presented in this section obtains analytic eigenenergies of a fundamental fermion under a linear potential using the present formalism \cite{Fuku84,Fuku14} mentioned in Section \ref{sec:Sec2}.
Our method expresses the fermion field in terms of the step-function-type basis functions localized in the spacetime continuum with finite degrees of freedom for a continuum limit.
The secular equation in the Hamiltonian matrix form corresponding to the classical Hamiltonian is analytically diagonalized, giving the lowest eigenvalue.
The squared system energies, with the large rotational energy compared to the composite fermion mass, are proportional to the string tension and relativistic quantum number of the angular momentum, which stem from the secular equation structure. We note that the constant of the angular momentum in the classical solution is explicitly derived in the relativistic quantum form.
This section first summarizes the relativistic classical Hamiltonian approach. We then give the corresponding relativistic Dirac equation, and the Dirac 
field is expressed in terms of the step-function type basis functions 
localized in the spacetime continuum, as mentioned in the previous 
section. Variational calculus leads to the secular equation, yielding the analytical eigenenergies. 
\par
The primary property of the Regge trajectory is described with the following relativistic classical mechanics. The theoretical method \cite{Miya, Maki} using the classical mechanical Hamiltonian in spherical coordinates is briefly summarized below.
The classical Hamiltonian $H^{({\rm cl})}$ is composed of the kinetic energy and linear potential given by
\begin{eqnarray}
\label{eqn:EMFM} 
H^{({\rm cl})}=(P^2+m^2)^{1/2}+\sigma r,
\end{eqnarray}
where $P$ is the relativistic momentum of a comprising fundamental fermion particle, $m$ is the fermion mass and $\sigma$ is the string tension between the fermion and antifermion.
Using the rotational quantity written by $J=Pr$, the above classical mechanical Hamiltonian, for the small mass compared to the kinetic (rotational) energy, is reduced to
\begin{eqnarray}
\label{eqn:EMD}
H^{(\rm cl)}=\frac{J}{r}+\sigma r.
\end{eqnarray}
The Hamiltonian $H^{(\rm cl)}$ takes the minimum energy $E^{(\rm cl) }_{\rm min}$ at $r=(J/\sigma)^{1/2}$, leading to
\begin{eqnarray}
\label{eqn:EMTO}
(E^{(\rm cl)}_{\rm min})^2=4 J\sigma.
\end{eqnarray}
The above relation is essentially consistent with the principal properties of the Regge trajectory \cite{ColReg}. 
\par

\subsection{Theoretical Dirac equation for a fermion-antifermion pair under a linear potential in spherical coordinates \label{sec:Sec62}}

We write a set of relativistic radial wave functions in the radial $r$-axis of spherical coordinates in the form
\begin{eqnarray}
\label{eqn:DWFN}
\psi_{F}(r)=\frac{F(r)}{r},
\hspace{2ex}
\psi_{G}(r)=\frac{G(r)}{r}.
\end{eqnarray}
Then, the total Hamiltonian in natural units
is given by
\begin{eqnarray}
\label{eqn:TOTH}
\nonumber
{\cal H}=\frac{1}{2}\int dr [F(+m-\frac{\alpha}{r})F
-F\frac{{\it d} G}{{\it d} r}-F\frac{\kappa}{r}G
\end{eqnarray}
\begin{eqnarray}
\nonumber
+ G(-m-\frac{\alpha}{r})G+G\frac{{\it d} F}{{\it d} r}-G\frac{\kappa}{r}F
\end{eqnarray}
\begin{eqnarray}
-FEF-GEG].
\end{eqnarray}
where $m$ is the mass of the fermion with the associated energy $E$, and 
$\alpha=g^2/(4\pi)$ with a coupling constant $g$.
The symbol $\kappa$ refers to the relativistic quantum number for the total angular momentum \cite{Schf}
with the following relation to the Dirac's notation $j_{\rm D}$ \cite{Dira}
\begin{eqnarray}
\kappa=j_{\rm D}=
\left\{\begin{array}{l}
l+1 \hspace{4ex}\mbox{for $j=l+1/2$ }\\
\mbox{ }\\
-l \hspace{6ex}\mbox{for $j=l-1/2$ }
\end{array}\right. ,
\end{eqnarray}
where $l$ is the quantum number for angular momentum.
Variational calculus leads to the following Dirac equations
\begin{eqnarray}
\label{eqn:DBSE}
(+m-\frac{\alpha}{r})F-\frac{{\it d} G}{{\it d} r}-\frac{\kappa}{r}G=EF,
\end{eqnarray}
\begin{eqnarray}
(-m-\frac{\alpha}{r})G+\frac{{\it d} F}{{\it d} r}-\frac{\kappa}{r}F=EG.
\end{eqnarray}
\par
For the above equations we consistently add the linear potential, derived from the Wilson loop from the non-Abelian gauge field, to the Coulomb potential, considering that energy is one component of the four-vector momentum.
Moreover, in the above equations, we  consider the case that
\begin{eqnarray}
|F\frac{{\it d} G}{{\it d} r}| <<| F\frac{\kappa}{r}G|,
\hspace{4ex}\mbox{that is,}\hspace{4ex}
|\frac{{\it d} G}{{\it d} r}| << |\frac{\kappa}{r}G|,
\end{eqnarray}
\begin{eqnarray}
|G\frac{{\it d} F}{{\it d} r}| << |G\frac{\kappa}{r}F|,
\hspace{4ex}\mbox{that is,}\hspace{4ex}
|\frac{{\it d} F}{{\it d} r}| <<| \frac{\kappa}{r}F|.
\end{eqnarray}
The above approximation corresponds to that the angular momentum, proportional to $\kappa$, is larger than the radial momentum, corresponding to the classical Hamiltonian. Then, Eq. (\ref{eqn:TOTH}) can be rewritten as
\begin{eqnarray}
\label{DPOTr}
\nonumber
{\cal H}=\frac{1}{2}\int dr [F(+m-\frac{\alpha}{r}+\sigma r)F
-F\frac{\kappa}{r}G
\end{eqnarray}
\begin{eqnarray}
\nonumber
                    + G(-m-\frac{\alpha}{r}+\sigma r)G
                    -G\frac{\kappa}{r}F
\end{eqnarray}
\begin{eqnarray}
-FEF-GEG].
\end{eqnarray}

\subsection{Dirac equation under linear potential expressed in terms of localized step-function-type basis functions in spacetime continuum \label{sec:Sec62}}

Similarly with orthogonal coordinates, we introduce lattice (grid) points $r_{n}$ $(n=1, 2, ....,N_{r})$ in the radial $r$-axis.
In terms of the step-function-type basis functions in Eq. (\ref{eqn:DeEBR}), we express the fermion wave functions in Eq. (\ref{eqn:DWFN}) in the form
\begin{eqnarray}
\label{SolF}
F(r)=\sum_{n}F_{n}\Omega^{\tilde{\rm E}}_{n}(r),
\end{eqnarray}
\begin{eqnarray}
\label{SolG} 
G(r)=\sum_{n}G_{n}\Omega^{\tilde{\rm E}}_{n}(r).
\end{eqnarray}
Then, it follows that
\begin{eqnarray}
\nonumber
\int dr F(r)G(r)
=\int dr \sum_{n}\sum_{k}
[F_{n}\Omega^{\tilde{\rm E}}_{n}(r) G_{k}\Omega^{\tilde{\rm E}}_{k}(r)]
\end{eqnarray}
\begin{eqnarray}
\nonumber
=\sum_{n}\sum_{k}[F_{n}G_{k}
\int dr 
\Omega^{\tilde{\rm E}}_{n}(r) \Omega^{\tilde{\rm E}}_{k}(r)]
=\sum_{n}\sum_{k}
(F_{n}G_{k}\Delta\delta_{n,k})
\end{eqnarray}
\begin{eqnarray}
=\Delta\sum_{n} F_{n}G_{n}.
\end{eqnarray}
Owing to the above relation, the aforementioned total Hamiltonian of the fermion-antifermion pair in Eq. (\ref{DPOTr}) is given by
\begin{eqnarray}
\label{eqn:HDIRD}
\nonumber
{\cal H}=\frac{\Delta}{2}
\nonumber
\sum_{n}
\{ [F_{n}(+m-\frac{\alpha}{r_{n}}+\sigma r_{n})F_{n}
-F_{n}\frac{\kappa}{r_{n}}G_{n}]
\end{eqnarray}
\begin{eqnarray}
+[G_{n}(-m-\frac{\alpha}{r_{n}}+\sigma r_{n})G_{n}
-G_{n}\frac{\kappa}{r_{n}}F_{n}]
-F_{n}EF_{n}-G_{n}EG_{n} \},
\end{eqnarray}
where
\begin{eqnarray}
\frac{\alpha}{r_{n}}=\frac{\alpha}{n\Delta},
\hspace{4ex}
\frac{\kappa}{r_{n}}=\frac{\kappa}{n\Delta}.
\end{eqnarray}
\par
Variational calculus imposes the normalization condition on $F$ and $G$, by replacing the last terms $-F_{n}EF_{n}-G_{n}EG_{n}$ in Eq. (\ref{eqn:HDIRD}) by $-E(F_{n}F_{n}-1/N_r)-E(G_{n}G_{n}-1/N_r)$, where $E$ acts as a Lagrange multiplier in this condition ($N_r$ is the number of lattice (grid) points in the radial $r$-axis).
By variational calculus, we obtain the following Dirac equation in matrix form for the $v$-th eigenvector composed of ${\bf F}^{v}$ and ${\bf G}^{v}$
with the associated eigenenergy $E^{v}$
\begin{eqnarray}
H
\left[\begin{array}{c}
{\bf F}^{v} \\
{\bf G}^{v} 
\end{array}\right]
=E^{v}
\left[\begin{array}{c}
{\bf F}^{v} \\
{\bf G}^{v} 
\end{array}\right],
\end{eqnarray}
where the components of the row vector
${\bf F}^{v}$ 
and ${\bf G}^{v}$
are $F^{v}_{i}$ and $G^{v}_{i}$, respectively, with $ 1 \leq i \leq N_r $.
(The notation
$i$ is not the imaginary unit in complex numbers, but an integer.)
The matrix has the form of
\begin{eqnarray}
\label{eqn:DHMt01}
H
=
\left[\begin{array}{cc}
{H}_{A} &{H}_{B} \\
{H}_{C} &{H}_{D} 
\end{array}\right]
=
\left[\begin{array}{cccccccc}
 + &0 &0 &0 &+ &0 &0 &0 \\
 0 &+ &0 &0 &0 &+ &0 &0 \\
 0 &0 &+ &0 &0 &0 &+ &0 \\
 0 &0 &0 &+ &0 &0 &0 &+ \\
 + &0 &0 &0 &+ &0 &0 &0 \\
 0 &+ &0 &0 &0 &+ &0 &0 \\
 0 &0 &+ &0 &0 &0 &+ &0 \\
 0 &0 &0 &+ &0 &0 &0 &+ \\
\end{array}\right],
\end{eqnarray}
where the matrix elements for $ 1 \leq i, j \leq N_r $ are
\begin{eqnarray}
\hspace{-8ex}
H_{Aij}=H_{ij}\delta_{ij}=(+m-\frac{\alpha}{i\Delta}
+\sigma i\Delta)
\delta_{ij},
\end{eqnarray}
\begin{eqnarray}
H_{Bij}=H_{i,N_r+j}=
\left\{\begin{array}{l}
\frac{-\kappa}{i\Delta} \hspace{4ex}\mbox{for $j=i$} \\
\mbox{ } \\
0 \hspace{4ex}\mbox{for the others} 
\end{array}\right. , 
\end{eqnarray}
\begin{eqnarray}
H_{Cij}=H_{N_r+i,j}=
\left\{\begin{array}{l}
\frac{-\kappa}{i\Delta} \hspace{4ex}\mbox{for $j=i$} \\
\mbox{ } \\
0 \hspace{4ex}\mbox{for the others} 
\end{array}\right. , 
\end{eqnarray}
\begin{eqnarray}
\label{eqn:DHMt20}
H_{Dij}=
H_{N_r+i,N_r+j}
\delta_{ij}=(-m-\frac{\alpha}{i\Delta}
+\sigma i\Delta)
\delta_{ij}.
\end{eqnarray}

\subsection{
Analytical derivation of eigenenergies for a fermion under linear potential with finite degrees of freedom for continuum limit \label{sec:Sec63}}

Here, we analytically calculate quantum eigenenergies recognizing that the energies of a classical mechanical Hamiltonian provide the principal properties of the Regge trajectory. 
From  the four elements $H_{ii}$, $H_{i,N_r+1}$, $H_{N_r+i,i}$ and $H_{N_r+i,N_r+i}$ of the Hamiltonian matrix in Eqs. (\ref{eqn:DHMt01})-(\ref{eqn:DHMt20}), a $2 \times 2$ sub-matrix can be constructed  without the relationship to the other elements. We diagonalize this sub-matrix by a unitary transformation with the matrix
\begin{eqnarray}
U_{pq}=
\left\{\begin{array}{c}
u_{pq} \hspace{4ex}\mbox{for $p, q=i, N_r+i$} \\
\mbox{ } \\
\hspace{-4ex} \delta_{pq} \hspace{4ex}\mbox{for the others} 
\end{array}\right.,
\end{eqnarray}
which has four elements $u_{pq}$ (in rank two) distinguished from the other elements. From the following determinant of the above sub-matrix,
\begin{eqnarray}
(E-H_{ii})
(E-H_{N+i,N+i})-H_{i,N+i}H_{N+i,i}
=0,
\end{eqnarray}
and disregarding the fermion masses, which is sufficiently small compared to the large rotational energy, we obtain
\begin{eqnarray}
\label{EDET}
E=-\frac{\alpha}{i\Delta}+\sigma i \Delta + \frac{|\kappa|}{i\Delta }.
\end{eqnarray}
(We note that $i$ is not a complex number but an integer of the lattice index).
In the above equation, the Coulomb term $-\alpha/(i\Delta)$ is dropped for sufficiently small $\Delta$ (in some sense beyond the regime of computer simulations with larger $\Delta$) considering the asymptotic freedom of the non-Abelian gauge field. Therefore, it follows that
\begin{eqnarray}
 \label{EqE}
E= \sigma i \Delta + \frac{|\kappa|}{i\Delta }.
\end{eqnarray}
\par
Here, we obtain the continuum limit of the above solution. Let $E_{x}$ be the following function of a real continuum number $x$ with $x > 0$
\begin{eqnarray}
E_{x}=\frac{|k|}{x}+\sigma x=(\sigma x+\frac{|k|}{x}).
\end{eqnarray}
This $E_{x}$ takes the minimum
\begin{eqnarray}
E_{x}^{\rm min}=2(|\kappa|\sigma)^{1/2},
\end{eqnarray}
at
\begin{eqnarray}
x_{\rm m}=(|\kappa|/\sigma)^{1/2}.
\end{eqnarray}
We measure the $x_{\rm m}$ with the lattice spacing $\Delta$, by expressing $x_{\rm m}$ as
\begin{eqnarray}
x_{\rm m}=i_{\rm m}\Delta+\epsilon_{\rm E},
\end{eqnarray}
where $i_{\rm m}$ represents a quantity corresponding to the lattice index (integer) and $\epsilon_{\rm E}$ is a residual denoted as
\begin{eqnarray}
-\frac{\Delta}{2} \leq \epsilon_{\rm E} < \frac{\Delta}{2}.
\end{eqnarray}
In the limit of $\Delta \rightarrow 0$, the residual $\epsilon_{\rm E}$ approaches $0$, and 
the eigenenergy $E$ in Eq. (\ref{EqE}) takes the minimum value $E_{\rm min}$, at a lattice point $i=i_{m}$, which coincides with $E_{x}^{\rm min}$, resulting in
\begin{eqnarray}
 \label{eqn:ESq}
E_{\rm min}^{2}=(E_{x}^{\rm min})^2
=4 |\kappa|\sigma,
\end{eqnarray}
which is independent of the lattice spacing $\Delta$.
\par
Here, we summarize the physical implication of the obtained results. 
The above Eq. (\ref{eqn:ESq}), which states that the squared energies (masses) are proportional to the absolute relativistic quantum number $|\kappa| $ of the total angular momentum, essentially reproduces the aforementioned classical result of Eq. (\ref{eqn:EMTO}), indicating the principal property of the Regge trajectory \cite{ColReg}. Experimentally, the slope in natural units observed from the Regge trajectory \cite{ColReg, Naga}, is
\begin{eqnarray}
\label{eqn:ESq2}
\frac{d|\kappa|}
{d (E_{\rm min}^2)}=0.93\hspace{1ex}[{\rm GeV^{-2}}].
\end{eqnarray}
From Eqs. (\ref{eqn:ESq})-(\ref{eqn:ESq2}), we obtain $\sqrt{\sigma}=518.5$ MeV, and, if we use the relation $\sqrt{\sigma}=2.255\Lambda_{\rm MOM}$, derived analytically in Eq. (\ref{eqn:StSIE}), with $\Lambda_{\rm MOM}$ being the scale-invariant energy of quantum chromodynamics (QCD), we have $\Lambda_{\rm MOM}=229.9$ MeV. This value of the scale-invariant energy is consistent with the experimental energy of around 213 MeV for QCD \cite{MPes}.
\par

\subsection{Relationship of relativistic quantum mechanical results and field-theoretical approach \label{sec:Sec64}}

Finally, the relativistic quantum results obtained in this section for a fermion-antifermion pair under a linear potential are related to the field-theoretical approach in the quenched case. This quantum quenched process is due to the Okubo-Zweig-Iizuka (OZI) rule \cite{Okub,Zwei,Iizu}, which implies the suppression of further fermion-antifermion pair creation.
For this process, the path integral with respect to the fermion Grassmann numbers $\bar{\Psi}$ and $\Psi$, containing a matrix $M_{\rm f}$ is represented as
\begin{eqnarray}
\int d \bar{\Psi} d\Psi \exp(- \bar{\Psi}M_{\rm f}\Psi)
={\rm det}(M_{\rm f}) \approx 1.
\end{eqnarray}
In the operator formalism, the Green's (two-point correlation) function at Euclidean time $t$ has the form
\begin{eqnarray}
\label{eqn:GreF}
G_{t0}(t)=<0|\hat{\cal{H}}_{\rm B}(t)
{\hat{\cal{H}}}^{\dag}_{\rm B}(0)|0>,
\end{eqnarray}
where $<0|$ and $|0>$ represent the ground-state vacuum. The symbol $\hat{\cal{H}}_{\rm B}$ is the Heisenberg-type Hamiltonian operator of the bound state (bound-state field), which is denoted, using the energy operator $\hat{E}$, as
\begin{eqnarray}
\hat{\cal{H}}_{\rm B}(t)=\exp(\hat{E}t)\hat{\cal{H}}_{\rm B}(0)\exp(-\hat{E}t).
\end{eqnarray}
From Eq. (\ref{eqn:DWFN}), the $s$-th solution of the Dirac equation $|s>$ with eigenenergy $E_{(s)}$ is given by
\begin{eqnarray}
|s> \propto
\left[\begin{array}{c}
\frac{G_{(s)}}{r} \\
            \\
\frac{F_{(s)}}{r}
\end{array}\right].
\end{eqnarray}
The above radial functions are multiplied by the spin-angular components and the total wave function in the center of mass coordinates. For the quenched case, the Green's function given by Eq. (\ref{eqn:GreF}) and the above solution $|s>$ yield
\begin{eqnarray}
\nonumber
G_{t0}(t)=\sum_s <0|\hat{\cal{H}}_{\rm B}(t)|s><s|
{\hat{\cal{H}}}^{\dag}_{\rm B}(0)|0>
\end{eqnarray}
\begin{eqnarray}
=\sum_s|<0|\hat{\cal{H}}_{\rm B}(0)|s>|^2
\exp(-E_{(s)} t).
\end{eqnarray}
Therefore, the eigenenergy of the Dirac equation derived by relativistic quantum mechanics has emerged as a decay constant for the Euclidean time in the field-theoretic formalism.
\par
Furthermore, the Polyakov line in the field-theoretic approach described in Subsection \ref{sec:Sec43} reveals the deconfinement of the fermion-antifermion pair at high temperatures.
The classical mechanical Hamiltonian shown in Subsection \ref{sec:Sec61} at absolute zero (temperature) does not describe the high-temperature phenomena.
From the Polyakov line, which we analytically derived, the binding energy $\epsilon_{\rm q}$ of the fundamental fermion and antifermion pair is given by Eq. (\ref{eqn:EPOLA}) as $\epsilon_{\rm q}=(\sigma r)/(k_{\rm B}T)$, where $k_{B}$ and $T$ are the Boltzmann constant and temperature, respectively. This expression indicates that the binding energy $\epsilon_{\rm q}$ is small at high temperatures and the deconfinement of the paired fermions occurs in some sense. This deconfinement is different from that for the squeezing of the electric flux tube by the superconductor because electric-flux-tube squeezing does not occur above the superconducting critical temperature.
\par
Consequently, owing to the OZI rule, which states that the further fermion-pair creation beyond one loop by the pair is suppressed, the solutions of the Dirac equation and its eigenenergies are consistently involved by the operator formalism. As shown in Subsection \ref{sec:Sec63}, the Dirac equation under a confining linear potential gives the lowest masses, which reproduce the Regge trajectory that the squared masses are proportional to the relativistic quantum number of the total angular momentum, when the rotational energy is larger than the masses of the composite fermion pair.
Moreover, owing to the interaction of the sources (particles) mediated by the non-Abelian gauge field, the Polyakov line describes the deconfinement at high temperatures in some sense, as shown in
Eq. (\ref{eqn:PolyL}).

\section{Further outlook \label{sec:Sec7}}

In this section, we describe the possibility of using the (Poincar${\rm \acute{e}}$ covariant) formalism in terms of the step-function-type basis functions presented in this paper for gravity. This formalism may be regarded as an approximation of the (quantum) field theory. However, one of the motivations for the development of the present theory is to construct a consistent (quantum field) theory, using the variational and path-integral methods involving gravity without the ultraviolet divergence. In the gravitational case, the basis differential equation of the conventional continuum theory is the Einstein equation, given by
\begin{eqnarray}
R_{{\rm g}\mu\nu}-\frac{1}{2}R_{\rm g}g_{\mu\nu}+\Lambda_{\rm g} g_{\mu\nu}
=\kappa_{\rm g}T_{{\rm g}\mu\nu},
\end{eqnarray}
where $R_{{\rm g}\mu\nu}$ is the curvature tensor of spacetime, $g_{\mu\nu}$ is the metric tensor, $\Lambda_{\rm g}$ is the cosmological constant, $\kappa_{\rm g}$ is the Einstein constant and $T_{{\rm g}\mu\nu}$ is the energy-momentum (stress-energy) tensor.
\par
The above equation has ultraviolet divergences, particularly at the quantum level. By expressing the gravitational field in terms of the step-function-type basis functions presented in this paper, the ultraviolet divergences are avoided.
At
(or around)
the
Planck
scale,
we expect that the step-function-type basis functions are suitable to describe the field compared to the conventional continuum theory because of the ultraviolet divergence in the latter theory.
Then, the difference in the energy obtained by the conventional continuum theory and the relatively realistic energy obtained using the step-function-type basis functions would compensate the energy of the continuum theory by involving the energy difference into the cosmological constant
(renormalizing the cosmological constant in some extended sense)
in view of the realistic theory.
(At the
Planck
scale, the realistic energy obtained using the step-function-type basis functions is approximated by the conventional continuum theory).
The order of magnitude of this energy difference would naturally be the same order of magnitude of the matter (atoms) \cite{Barrow}, and the above energy difference
may be regarded as dark energy.
(An example for dark matter may be the vacuum energy
of the soliton-like objects considered in Section \ref{sec:Sec4} of
this paper.) 
This cosmological constant obtained from the above energy difference would be relatively tiny compared to the huge value expected from some conventional theories.
Since the continuum theory overestimates the field (such as a simple example proportional to $1/r$) and the spacetime curvature near the center of a particle compared to the step-function-type wave basis functions, the stress by the cosmological constant would be repulsive. (Even if the initial energy density is uniform, some energy gradient would arise by the energy propagation before inflation.) In a small region with a high energy density, the above energy difference
owing to the difference in the wave function
is huge, and could lead to inflation (expansion) of the early universe, followed by relaxation toward the 
decrease in the density of the energy difference
due to the difference in the wave function of fields expressing such as curvatures.
Therefore, as mentioned above, it may be useful to use the present step-function-type basis function for gravitational research.

\section{Conclusions \label{sec:Sec8}}

In this article, we have reviewed the authors' research concerning the construction of a consistent Poincar${\rm \acute{e}}$ covariant field-theoretic formalism in terms of step-function-type basis functions without ultraviolet divergences for non-pure/pure non-Abelian Yang-Mills gauge fields. By using this formalism, we analytically derived the characteristics for the confinement/deconfinement, mass-gap and Regge trajectory for non-Abelian gauge fields and obtained self-consistent solutions for the self-interacting scalar fields.
In the present formalism, fields are expressed in terms of the step-function-type basis functions with finite degrees of freedom toward the continuum limit in a parameter spacetime continuum mapped to real spacetime continuum. The present formalism is Poincar${\rm \acute{e}}$ covariant and gauge invariant without ultraviolet divergences, showing that the 
consistent non-pure/pure
non-Abelian Yang-Mills gauge field generated from group theory is defined to exist, which is the requirement from fundamental theoretical physics.
We have derived a new classical solution of the non-Abelian Yang-Mills equation for a fundamental particle-antiparticle pair, which reveals the classical confining linear potential caused by the trace of the polynomials of the Lie matrices associated with the confining soliton-like field configuration. It was derived that the quantum action has the local mass, and the Polyakov line indicates finite binding energies (masses) at low temperatures. This mechanism also works for the pure non-Abelian Yang-Mills field because the source and antisources are created.
These confinement properties indicate that non-pure/pure non-Abelian Yang-Mills fields have the positive mass-gap, which is desired to satisfy Pauli's requirement.
Furthermore, solutions of the Dirac equation under a linear potential were analytically derived in this formalism for small masses of a fermion and antifermion compared to the rotational energy, and reproduced the principal properties of Regge trajectories at a quantum level. Squared masses (energies) of the system are a function of the string tension and relativistic quantum number of angular momentum, which dose not appear at the classical mechanical level.
In addition, (in Further outlook) we have mentioned the possibility that the cosmological constant may be
caused by
the energy difference due to the difference between the wave function (of fields) of the conventional continuum theory and the wave function (of fields) in terms of the step-function-type basis functions expressing such as curvatures.
\par




\end{document}